\documentclass[12pt]{article}

\usepackage[utf8]{inputenc}

\usepackage[sectionbib]{natbib}
\usepackage{array,epsfig,rotating}
\usepackage{color}
\usepackage[dvipsnames]{xcolor}
\usepackage{hyperref}
\hypersetup{
  colorlinks,
  linkcolor={red!50!black},
  citecolor={blue},
  urlcolor={blue!80!black}
}

\usepackage{tabularx}
\usepackage{amsmath}
\usepackage{amssymb}
\usepackage{amsfonts}
\usepackage{amsthm}
\usepackage{float}
\usepackage{verbatim}
\usepackage{setspace}
\usepackage{caption}
 \usepackage{subfigure}
\usepackage{algorithm}
\usepackage{algpseudocode}
\usepackage{multirow}
\usepackage{booktabs}
\usepackage[title]{appendix}

\algrenewcommand\algorithmicrequire{\textbf{Input:}}
\algrenewcommand\algorithmicensure{\textbf{Output:}}

\newtheorem{theorem}{Theorem}

\newtheorem{condition}{Condition}
\newtheorem{lemma}{Lemma}

\newtheorem{proposition}{Proposition}
\newtheorem{definition}{Definition}
\newtheorem{example}{Example}
\newtheorem{remark}{Remark}

\allowdisplaybreaks

\newcommand{\bo}{\boldsymbol }

\newcommand{\one}{\mathbf 1}

\newcommand{\rank}{\text{\normalfont{rank}}}

\newcommand{\A}{\mathbf A}
\newcommand{\B}{\mathbf B}
\newcommand{\C}{\mathbf C}

\newcommand{\e}{\mathbf e}

\newcommand{\bH}{\mathbf H}

\newcommand{\I}{\mathbf I}

\newcommand{\M}{\mathbf M}

\def\S{\mathbf S}
\newcommand{\bO}{\mathbf O}
\newcommand{\bP}{\mathbf P}

\newcommand{\Q}{\mathbf Q}
\newcommand{\R}{\mathbf R}

\newcommand{\U}{\mathbf U}

\newcommand{\W}{\mathbf W}
\newcommand{\V}{\mathbf V}

\newcommand{\x}{\mathbf x}
\newcommand{\Y}{\mathbf Y}

\newcommand{\bPi}{\boldsymbol{\Pi}}

\newcommand{\bpi}{\boldsymbol{\pi}}
\newcommand{\bT}{\boldsymbol{\Theta}}
\newcommand{\bt}{\boldsymbol{\theta}}
\newcommand{\bS}{\boldsymbol{\Sigma}}

\DeclareMathOperator*{\argmax}{argmax} 

\def\hat{\widehat}
\def\tilde{\widetilde}

\newcommand{\zero}{\mathbf 0}

\newcommand{\norm}[1]{\left\lVert#1\right\rVert}

\usepackage[margin=1in]{geometry}

\newcommand\blfootnote[1]{%
  \begingroup
  \renewcommand\thefootnote{}\footnote{#1}%
  \addtocounter{footnote}{-1}%
  \endgroup
}

\usepackage{mathtools}

\def\spacingset#1{\renewcommand{\baselinestretch}%
{#1}\small\normalsize} 
\spacingset{1}

\theoremstyle{definition}

\title{A Spectral Method for Identifiable Grade of Membership Analysis with Binary Responses}
\author{\large Ling Chen$^*$ and Yuqi Gu$^\dagger$\\[4mm]
Department of Statistics, Columbia University}
\date{}

\begin{document}

\maketitle

\begin{abstract}
    Grade of Membership (GoM) models are popular individual-level mixture 
    models for multivariate categorical data. GoM allows each subject to have mixed memberships in multiple extreme latent profiles. Therefore GoM models have a richer modeling capacity than latent class models that restrict each subject to belong to a single profile.
    The flexibility of GoM comes at the cost of more challenging identifiability and estimation problems. In this work, we propose a singular value decomposition (SVD) based spectral approach to GoM analysis with multivariate binary responses. 
    Our approach hinges on the observation that  the expectation of the data matrix has a low-rank decomposition under a GoM model.
    For \emph{identifiability}, we develop sufficient and almost necessary conditions for a notion of expectation identifiability.
    For \emph{estimation}, we extract only a few leading singular vectors of the observed data matrix, and exploit the simplex geometry of these vectors to estimate the mixed membership scores and other parameters.
    We also establish the consistency of our estimator in the double-asymptotic regime where both the number of subjects and the number of items grow to infinity.
    Our spectral method has a huge computational advantage over Bayesian or likelihood-based methods and is scalable to large-scale and high-dimensional data. 
    Extensive simulation studies demonstrate the superior efficiency and accuracy of our method. We also illustrate our method by applying it to a personality test dataset.
\end{abstract}

\noindent\textbf{Keywords}:
Grade of Membership Model;
Identifiability; 
Latent Variable Model;
Mixed Membership Model; 
Sequential Projection Algorithm;
Singular Value Decomposition; 
Spectral Method.

\blfootnote{$^*$Email: \texttt{lc3521@columbia.edu}.}
\blfootnote{$^\dagger$Email: \texttt{yuqi.gu@columbia.edu}. This work is partially supported by NSF Grant DMS-2210796.}

\spacingset{1.5}

\section{Introduction}
Multivariate categorical data are routinely collected in various social and behavioral sciences, such as 
psychological tests \citep{chen2019joint},
educational assessments \citep{shang2021partial}, and
political surveys \citep{chen2021unfolding}.
In these applications, it is often of great interest to use latent variables to model the unobserved constructs such as personalities, abilities, political ideologies, etc.
Popular latent variable models for multivariate categorical data include
the item response theory models \citep[IRT;][]{embretson2013item} and latent class models \citep[LCM;][]{hagenaars2002applied}, which employ continuous and discrete latent variables, respectively. 
Different from these modeling approaches, the grade of membership (GoM) models \citep{woodbury1978gom, erosheva2002grade, erosheva2005comparing} allow each observation to have mixed memberships in multiple \emph{extreme latent profiles}. 
GoM models assume that each observation has a latent membership vector with $K$ continuous membership scores that sum up to one. Each membership score quantifies the extent to which this observation belongs to each of $K$ extreme profiles. 
So GoM can be viewed as incorporating both the continuous aspect (via the membership scores) and discrete aspect (via the $K$ extreme latent profiles) of latent variables.
{More generally, GoM belongs to the broad family of mixed membership models for individual-level mixtures \citep{airoldi2014handbook}. Thanks to their nice interpretability and rich expressive power, variants of mixed membership models including GoM are widely used in many applications such as survey data modeling \citep{erosheva2007aoas}, response time modeling \citep{pokropek2016time}, topic modeling \citep{blei2003lda},
social networks \citep{airoldi2008mmsbm}, 
and data privacy \citep{manrique2012jasa}.}

The flexibility of GoM models comes at the cost of  more challenging identifiability and estimation problems.
In the existing literature on GoM model estimation, Bayesian inference using Markov Chain Monte Carlo (MCMC) is perhaps the most prevailing approach  \citep{erosheva2002grade, erosheva2007aoas, gormley2009grade, gu2023dimension}. However, 
the posterior distributions of the GoM model parameters are complicated, after integrating out the individual-level membership scores.
Many studies developed advanced MCMC algorithms for approximate posterior computation.
Yet MCMC sampling is time-consuming and typically not computationally efficient.
On the other hand, the frequentist estimation approach of marginal maximum likelihood (MML)  would bring a similar challenge, because the marginal likelihood still involves the intractable integrals of those latent membership scores.
Actually, it was pointed out in \cite{borsboom2016kinds} that GoM models are very useful in identifying meaningful profiles in applications including depression, personality disorders, etc., but they are temporarily not widely used in psychometrics due to the lack of readily accessible and efficient statistical software.

Recently, the development of the R package \texttt{sirt} \citep{robitzsch2022package} provides a joint maximum likelihood (JML)  algorithm for GoMs based on the iterative estimation method proposed in \cite{erosheva2002grade}.
In contrast to MML, the JML approach treats the subjects' latent membership scores as fixed unknown parameters rather than random quantities. This approach hence circumvents the need of evaluating the intractable integrals during estimation.
JML is currently considered the most efficient tool for estimating GoM models. However, due to its iterative manner, JML's efficiency is still unsatisfactory when applied to very large-scale data with many observations and many items. Therefore, it is desirable to develop more scalable and non-iterative estimation methods and aid psychometric researchers and practitioners to perform GoM analysis of modern item response data.

In addition to the difficulty of estimation, model identifiability is also a challenging issue for GoM models.
A model is identifiable if the model parameters can be reliably recovered from the observed data.
Identifiability is crucial to ensuring valid statistical estimation  as well as meaningful interpretation of the inferred latent structures. 
The handbook of \cite{airoldi2014handbook} emphasizes theoretical difficulties of identifiability in mixed membership models, including GoM models.
Recently, recognizing the difficulty of establishing identifiability of GoM models, \cite{gu2023dimension} proposed to incorporate a dimension-grouping modeling component to GoM and established the population identifiability for this new model. However, their identifiability results do not apply to the original GoM. In addition, their identifiability notion only concerns the population parameters in the model, but excludes the individual-level latent membership scores.

To address the aforementioned issues, we propose a novel singular value decomposition (SVD) based spectral approach to GoM analysis with multivariate binary data. Our approach hinges on the observation that the expectation of the response matrix admits a low-rank decomposition under GoM. 
Our contributions are three-fold.
\emph{First}, we consider a notion of \emph{expectation identifiability} and establish identifiability for GoM models with binary responses. 
Under this new notion, the identifiable quantities include not only the population parameters, but also the individual membership scores that indicate the grades of memberships.
Specifically, we derive sufficient conditions that are almost necessary for identifiability. 
\emph{Second}, based on our new identifiability results, we propose an SVD-based {spectral} estimation method scalable to large-scale and high-dimensional data.
\emph{Third}, we establish the consistency of our spectral estimator in the double-asymptotic regime where both the number of subjects $N$ and the number of items $J$ grow to infinity. Both the population parameters and the individual membership scores can be consistently estimated on average.
In the simulation studies, we empirically verify the identifiability results and also demonstrate the superior efficiency and accuracy of our algorithm. A real data example also illustrates that meaningful interpretation can be drawn after applying our proposed method.

The rest of the paper is structured as follows. Section \ref{sec: model} introduces the model setup and lays out the motivation for this work. Section \ref{sec: identifiability} presents the identifiability results.  Section \ref{sec: algorithm} proposes a spectral estimation algorithm and establishes its consistency.  Section \ref{sec: simulation} conducts simulation studies to assess the performance of the proposed method and empirically verify the identifiability results.
Section \ref{sec: real} illustrates the proposed method using a real data example in a psychological test.
Finally, Section \ref{sec: discussion} concludes the paper and discusses future research directions.
The proofs of the identifiability results are included in the Appendix.

\section{Model Setup and Motivation} \label{sec: model}
GoM models can be used to model multivariate categorical data with a mixed membership structure.
In this work, we focus on multivariate binary responses, which are very commonly encountered in social, behavioral, and biomedical applications, including yes/no responses in social science surveys \citep{erosheva2007aoas, chen2021unfolding}, correct/wrong answers in educational assessments \citep{shang2021partial}, and presence/absence of symptoms in medical diagnosis \citep{woodbury1978gom}.
We point out that our identifiability results and spectral estimation method in the binary case will illuminate the key structure of GoM and pave the way for generalizing to the general categorical response case.
We will  briefly discuss the possibility of such extensions in Section \ref{sec: discussion}.
In our binary response setting, denote the number of items by $J$.
For a random subject $i$, denote his or her observed response to the $j$-th item by $R_{ij}\in\{0,1\}$ for  
$j=1,\dots, J$.


A GoM model is characterized by two levels of modeling: the population level and the individual level. On the population level, $K$ \emph{extreme latent profiles} are defined to capture a finite number of prototypical response patterns. For $k\in1, \dots, K$, the $k$-th extreme latent profile is characterized by the item parameter vector $\bt_{k}=(\theta_{1k},\dots, \theta_{JK})$ 
with $\theta_{jk}\in[0,1]$ collecting the Bernoulli parameters of conditional response probabilities. Specifically, 
\begin{align}\label{eq-theta}
    \theta_{jk} = \mathbb P(R_{ij}=1\mid \text{subject $i$ solely belongs to the $k$-th extreme latent profile}).
\end{align}
We collect all the item-level Bernoulli parameters in a $J\times K$ matrix $\bT=(\theta_{jk})\in\mathbb{R}^{J\times K}$. 
On the individual level, each subject $i$ has a latent membership vector $\bpi_i=(\pi_{i1},\dots,\pi_{iK})$, satisfying $\pi_{ik}\ge 0$ and $\sum_{k=1}^K \pi_{ik}=1$. Here $\pi_{ik}$ indicates the extent to which subject $i$ partially belongs to the $k$-th extreme profile, and they are called membership scores.
It is now instrumental to compare the assumption of the GoM model with that of the latent class model \citep[LCM;][]{goodman1974, hagenaars2002applied}.
Both GoM and LCM share a similar formulation of the item parameters $\bo\Theta$ as defined in \eqref{eq-theta}.
However, under an LCM, each subject $i$ is associated with a categorical variable $z_i\in[K]$ instead of a membership vector. This means LCM restricts each subject to solely belonging to a single profile, as opposed to partially belonging to multiple profiles in the GoM. 
Further, the fundamental representation theorem in \cite{erosheva2007aoas} shows that a GoM model can be reformulated with an LCM but with $K^J$ latent classes instead of $K$ latent classes.
In summary, GoM is a more general and flexible tool than LCM for modeling multivariate data, but also exhibits a more complicated model structure.
It is therefore more challenging to establish identifiability and perform estimation under the GoM setting.


Given the membership score vector $\bpi_i$ and the item parameters $\bo\Theta$, the conditional probability of the $i$-th subject providing a positive response to the $j$-th item is
\begin{equation}\label{eq-convex}
    \mathbb P(R_{ij}=1\mid\bpi_i, \bo\Theta) = \sum_{k=1}^K \pi_{ik}\theta_{jk}.
\end{equation}
In other words, the positive response probability for a subject to an item is a convex combination of the extreme profile response probabilities $\theta_{jk}$ weighted by the subject's membership scores $\pi_{ik}$. 
The GoM model assumes that a subject's responses $R_{i,1},\dots, R_{i, J}$ are conditionally independent given the membership scores $\bo\pi_i$.
We consider a sample of $N$ i.i.d. subjects, and collect all the membership scores in a $N\times K$ matrix $\bPi=(\pi_{ik})\in\mathbb{R}^{N\times K}$.

In the GoM modeling literature \citep[e.g.][]{erosheva2002grade}, there are two perspectives of dealing with the membership scores $\bpi_i$: the random-effect and the fixed-effect perspectives.
The random effect perspective treats $\bpi_i$ as random and assumes that they follow some distribution parameterized by $\boldsymbol\alpha$: $\bpi_i \sim D_{\boldsymbol\alpha}(\cdot)$. 
Note that $\bpi_i\in \Delta_{K-1} = \{\x=(x_1,\dots, x_K): x_i\ge 0,\ \sum_{i=1}^Kx_i=1\}$ where $\Delta_{K-1}$ denotes the probability simplex.
A common choice of the distribution $D_{\boldsymbol\alpha}$ for $\bo\pi_i$ is the Dirichlet distribution \citep{blei2003lda}.


From the above random-effect perspective,
the \emph{marginal likelihood} function for a GoM model is
\begin{equation}\label{eq-ml}
    L(\bT, \boldsymbol\alpha\mid \R) = \prod_{i=1}^N
    \int_{\Delta_{K-1}} \prod_{j=1}^J \left(\sum_{k=1}^K\pi_{ik}\theta_{jk}\right)^{R_{ij}}\left(1-\sum_{k=1}^K\pi_{ik}\theta_{jk}\right)^{1-R_{ij}} d D_{\boldsymbol\alpha}(\bpi_i),
\end{equation}
where the membership scores $\bpi_i$'s are marginalized out with respect to their distribution $D_{\bo\alpha}$.
The integration of the products of sums in \eqref{eq-ml} poses challenges to establishing identifiability. This difficulty motivated \cite{gu2023dimension} to introduce a dimension-grouping component to simplify the integrals and then prove identifiability for that new model. 
In terms of estimation, the marginal maximum likelihood (MML) approach maximizes \eqref{eq-ml} to estimate the population parameters $(\bo\Theta,\bo\alpha)$ rather than the individual membership scores $\bPi$.
Bayesian inference with MCMC is also often used for estimation \citep{erosheva2007aoas, manrique2012jasa, gu2023dimension}, where inferring parameters like $\bo\alpha$ in the Dirichlet distribution $D_{\bo\alpha}(\cdot)$ typically require the Metropolis-Hastings sampling.

On the other hand, the fixed-effect perspective of GoM treats the membership scores $\bo\pi_i$ as fixed unknown parameters and aims to directly estimate them.
This approach does not model the distribution of the membership scores, and hence circumvents the need to evaluate the intractable integrals during estimation.
In this case, if still adopting the likelihood framework, the \emph{joint likelihood} function of both $\bo\Theta$ and $\bo\Pi$ for a GoM model is 
\begin{equation}\label{eq-jl}
    L(\bPi, \bT \mid \R) = \prod_{i=1}^N\prod_{j=1}^J \left(\sum_{k=1}^K\pi_{ik}\theta_{jk}\right)^{R_{ij}}\left(1-\sum_{k=1}^K\pi_{ik}\theta_{jk}\right)^{1-R_{ij}}.
\end{equation}
The joint maximum likelihood (JML) approach maximizes \eqref{eq-jl} to estimate  $\bo\Theta$ and $\bo\Pi$.
Based on an iterative algorithm proposed by \cite{erosheva2002grade}, the R package \texttt{sirt} \citep{robitzsch2022package} provides a function \texttt{JML} to solve this optimization problem under GoM. JML methods are typically known as inconsistent for many traditional models \citep{neyman1948consistent} when the sample size goes to infinity (large $N$) but the number of observed variables is finite (fixed $J$).
Nonetheless, in modern large-scale assessments or surveys where the data collection scope is unprecedentedly big and high-dimensional, both $N$ and $J$ can be quite large. 
JML is currently considered the most efficient tool for estimating GoM models. However, due to its iterative manner, JML's efficiency is still unsatisfactory when applied to modern big datasets with many observations and many items. Therefore, it is desirable to develop more scalable and non-iterative estimation methods and aid psychometric researchers and practitioners in performing GoM analysis of item response data.

To address the above issues in the GoM analysis,  
in this work, we propose a novel singular value decomposition (SVD) based spectral approach. 
Our approach hinges on the observation that the expectation of the response matrix under a GoM model admits a low-rank decomposition.
To see this, it is useful to summarize \eqref{eq-convex} in matrix form 
\begin{equation}\label{eq-model}
    \R_0:=\mathbb{E}[\R]=\underbrace{\bPi}_{N\times K} \underbrace{\bT^{\top}}_{K\times J},
\end{equation}
where $\R=(R_{ij})\in\{0,1\}^{N\times J}$ denotes the binary response data matrix, and  $\R_0=(R_{0,ij})\in\mathbb{R}^{N\times J}$ is the element-wise expectation of $\R$. 
Note that the factorization in \eqref{eq-model} implies that the $N\times J$ matrix $\R_0$ has rank at most $K$, which is the number of extreme latent profiles. Since $K$ is typically (much) smaller than $N$ and $J$, 
the decomposition \eqref{eq-model} exhibits a low-rank structure.
Therefore, we can consider the singular value decomposition (SVD) of $\R_0$:
\begin{equation}\label{eq-svd}
    \R_0=\U\bS \V^{\top},
\end{equation}
where $\bS$ is a $K\times K$ diagonal matrix collecting the $K$ singular values of $\R_0$; denote these singular values by $\sigma_1\ge\dots \ge \sigma_K\ge 0$ and write $\bo\Sigma=\text{diag}(\sigma_1,\dots, \sigma_K)$. Matrices $\U_{N\times K}$, $\V_{J\times K}$ collect the corresponding left and right singular vectors and satisfy $\U^{\top}\U=\V^{\top}\V=\mathbf{I}_K$.
Our high-level idea is to utilize the top $K$ left singular vectors of the data matrix $\R$ to identify and estimate $\bo\Pi$ and subsequently, $\bo\Theta$.
In the following two sections, we will present new identifiability results and develop a spectral estimation algorithm for GoM models based on the SVD in \eqref{eq-svd}.

\section{Identifiability Results} \label{sec: identifiability}


The study of identifiability in statistics dates back to \cite{koopmans1950identification}. 
A model is identifiable if the model parameters can be reliably recovered from the observed data.
Identifiability is a crucial property of a statistical model as it is a prerequisite for valid and reproducible statistical inference. In latent variable modeling, identifiability is especially essential since it is a foundation for meaningful interpretation of the latent constructs.

Traditionally, identifiability of a statistical model means that the population parameters can be uniquely determined from the marginal distribution of the observed variables \citep{koopmans1950identification, goodman1974}. In the context of GoM models, this notion of \emph{population identifiability} is equivalent to identifying parameters $(\bo\Theta, \bo\alpha)$ from the marginal distribution in \eqref{eq-ml}.
The complicated integrals in \eqref{eq-ml} make it difficult to establish population identifiability, which motivated \cite{gu2023dimension} to propose a dimension-grouping modeling component to simplify GoM and prove identifiability for that new model.
However, it remains unknown whether the original GoM models can be identified.

In this work, we consider a new notion of identifiability, which we term as \emph{expectation identifiability}. This notion concerns not only the item parameters, but also the individual membership scores. 
Similar identifiability notions are widely adopted and studied in the network modeling and topic modeling literature, e.g., \cite{jin2023mixed}, \cite{ke2023special}, \cite{mao2021mm}, and \cite{ke2022svd}.
Specifically, recall from \eqref{eq-model} that the expectation of the data matrix $\R$ has a low-rank decomposition $\R_0 = \bo\Pi \bo\Theta^\top$, we seek to understand under what conditions this decomposition is unique. Note that both the Bernoulli probabilities $\bo\Theta$ and the membership scores $\bo\Pi$ are treated as parameters to be identified.
We call a parameter set $(\bPi,\bT)$ valid if $\bo\pi_i\in\Delta_{K-1}$ and $\theta_{jk}\in[0,1]$ for all $i$, $j$, and $k$.
We formally define expectation identifiability below.

%

\begin{definition}[Expectation identifiability]\label{def: identifiability}
    A GoM model with parameter set $(\bPi, \bT)$ is said to be identifiable, if for any other valid parameter set $(\tilde{\bPi}, \tilde{\bT})$, $\tilde{\bPi}\tilde{\bT}^\top=\bPi\bT^{\top}$ holds if and only if $(\bPi, \bT)$ and $(\tilde{\bPi}, \tilde{\bT})$ are identical up to a permutation of the $K$ extreme profiles.
\end{definition}


One might ask whether identifying $\bo\Pi$ and $\bo\Theta$ from the expectation $\R_0$ has any implications on identifying $\bo\Pi$ and $\bo\Theta$ from the observed data $\R$. 
In fact, when both $N$ and $J$ are large with respect to $K$, it is known that the difference between the low-rank decompositions of $\R_0$ and $\R$ is small in a certain sense \citep{chen2021spectral}. We will revisit and elaborate on this subtlety when describing our estimation method and presenting the simulation results.
%
%
Briefly speaking, studying the expectation identifiability problem from $\R_0$ is very meaningful in modern large-scale and high-dimensional data settings with large $N$ and $J$.

We next present our new identifiability conditions for GoM models.
We first define the important concept of pure subjects.

\begin{definition}[Pure subject]
    Subject $i$ is a pure subject for extreme profile $k$ if the only positive entry of $\bpi_i$ is located at index $k$; that is,
    $$
    \bpi_i = (0, \dots, 0, \underbrace{1}_{\text{$k$-th entry}}, 0, \dots, 0).
    $$
\end{definition}
In words, subject $i$ is a pure subject for profile $k$ if it solely belongs to this profile and has no membership in any other profile.
We consider the following condition for $\bo\Pi$.

\begin{condition}\label{condition1}
    $\bo\Pi$ satisfies that every extreme latent profile has at least one pure subject. 
\end{condition}

Condition~\ref{condition1} is a quite mild assumption on $\bo\Pi$, 
because it only requires that each of the $K$ extreme profiles has at least one representative subject among all $N$ subjects. Intuitively, this condition is reasonable because the existence of these representative subjects indeed helps pinpoint the meaning and interpretation of the extreme profiles.
 
In real data applications of the GoM model, each pure subject is characterized by a prototypical response pattern indicating a particular classification or diagnosis. 
Specifically, each column of the $J\times K$ item parameter matrix $\bT$ is examined to coin the interpretation of each extreme profile. So, the existence of a pure subject in the $k$th ($1\leq k\leq K$) extreme profile means that there indeed exists a prototypical subject characterized by the parameters in the $k$th column of the $\bT$ matrix.
As a concrete applied example, \cite{woodbury1978gom} fitted the GoM model to a clinical dataset, where the item parameters for four extreme latent profiles were estimated. Then each of the extreme profiles was interpreted according to its response characteristics revealed via the item parameters. 
There, the four extreme profiles were interpreted as ``Asymptomatic'', ``Moderate'', ``Acyanotic Severe'', ``Cyanotic Severe'' in \cite{woodbury1978gom}. 
In this context, having pure subjects in each of these extreme profiles is a practically meaningful assumption, because it just means that in the sample with a large number of $N$ subjects, there exist an ``Asymptomatic'' subject, a ``Moderate'' subject, an ``Acyanotic Severe'' subject, and a ``Cyanotic Severe'' subject.

Under Condition~\ref{condition1}, $\bPi$ contains one identity submatrix $\mathbf{I}_K$ after some row permutation. 
For any matrix $\A$, we use $\A_{\S,:}$ to denote the rows with indices in $\S$. 
Under Condition \ref{condition1}, denote $\S=(S_1,\dots, S_K)$ as the index vector of one set of $K$ pure subjects such that $\bPi_{\S,:}=\mathbf{I}_K$.
So $S_1, \dots, S_K$ are distinct integers ranging in $\{1,\dots, N\}$.
For example, if the first $K$ rows of $\bo\Pi$ is equal to $\I_K$, then $\S = (1,2,\dots,K)$.
Recall $\R_0 = \U\bo\Sigma\V^\top$ is the SVD for $\R_0$, where $\U$ is a $N\times K$ matrix collecting the $K$ left singular vectors as columns.
Interestingly, Condition~\ref{condition1} induces a \emph{simplex geometry} on the row vectors of $\U$.
We have the following important proposition, which serves as a foundation for both our identifiability results and estimation procedure.

\begin{proposition}\label{prop1}
    Under Condition~\ref{condition1}, the left singular matrix $\U$ satisfies
    \begin{equation}\label{eq-simplex}
    \U=\bPi\U_{\S,:}.
\end{equation}
Furthermore, $\bPi$ and $\bT$ can be written as
\begin{align} 
    \label{eq-pi_lemma1}
    \bPi &= \U \U_{\S,:}^{-1},\\
    \label{eq-theta_lemma1}
    \bT & = \V\bS\U^{\top}\bPi(\bPi^{\top}\bPi)^{-1} = \V\bo\Sigma \U^\top_{\S,:}.
\end{align}
\end{proposition}

\begin{figure}[hbt!]
    \centering
    \includegraphics[width=0.5\textwidth]{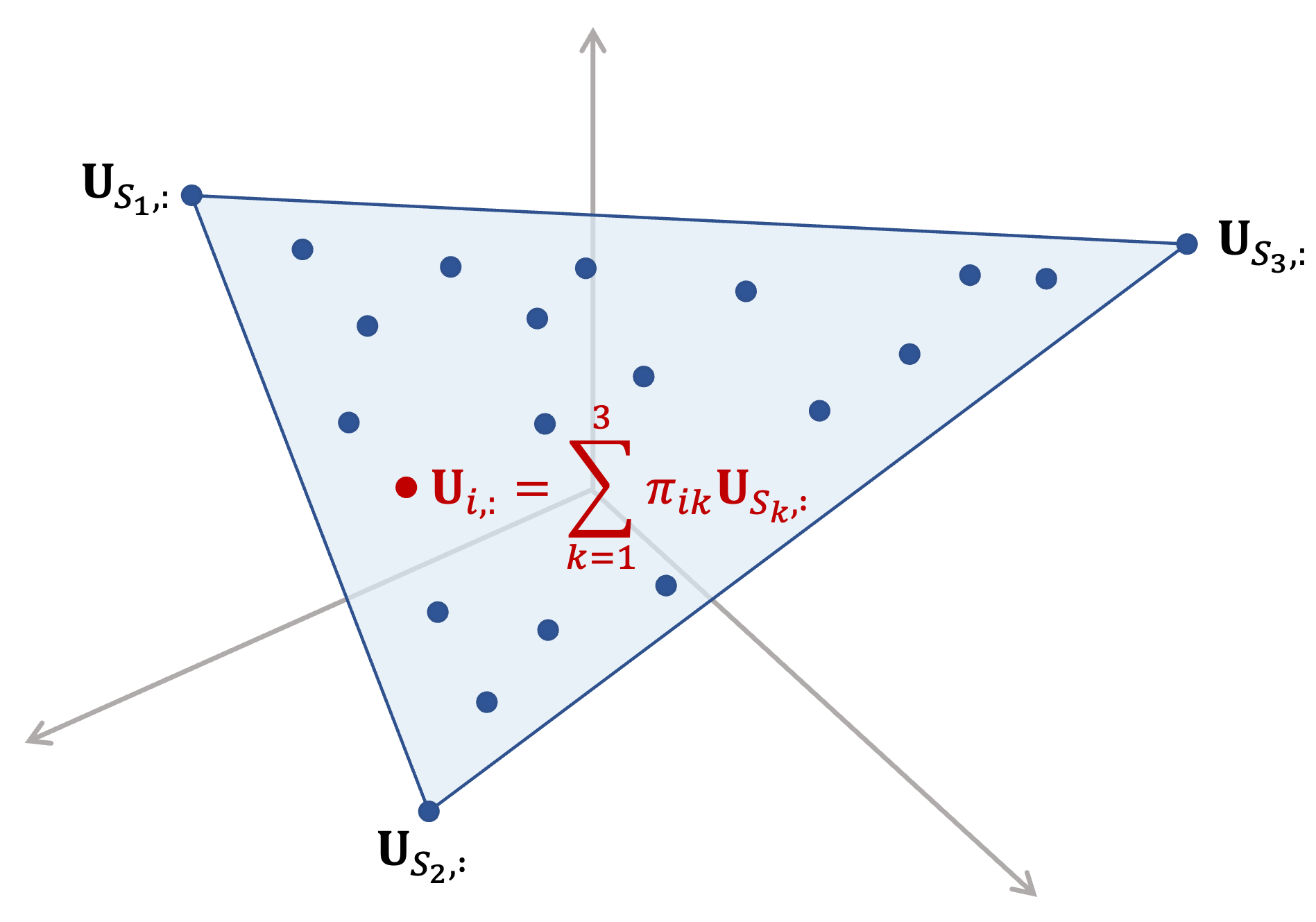}
    \caption{Illustration of the simplex geometry of the $N\times K$ left singular matrix $\U$ with $K=3$. The solid dots represent the row vectors of $\U$ in $\mathbb R^3$, and the three simplex vertices (i.e, vertices of the triangle) correspond to the three types of pure subjects. All the dots lie in this triangle.}
    \label{fig-simplex}
\end{figure}

We elaborate more on Proposition \ref{prop1} below.
Equation \eqref{eq-simplex} in Proposition \ref{prop1} implies that the left singular matrix $\U$ and the membership score matrix $\bo\Pi$ differ by a linear transformation, which is the $K\times K$ matrix $\U_{\mathbf S, :}$. 
Condition 1 and the properties of the singular value decomposition (such as the columns of $\mathbf V$ being orthogonal to each other and $\bS$ being invertible) are used to prove \eqref{eq-simplex}.
Equations \eqref{eq-pi_lemma1} and \eqref{eq-theta_lemma1} imply that if an index set of pure subjects $\S$ is known, then the parameters of interest $\bPi$ and $\bT$ can be written in closed forms in terms of the SVD and $\S$.
More specifically, \eqref{eq-simplex} is equivalent to
\begin{equation}
    \U_{i, :}
    =\sum_{k=1}^K\pi_{ik}\U_{S_k, :}, \ i=1,\dots, N.
\end{equation}
Each $\U_{i,:}$ is the embedding of the $i$-th subject into the top-$K$ left singular subspace of $\R_0$, and all the rows $\U_{1,:}, \ldots, \U_{N,:}$ can be plotted as points in the $K$-dimensional Euclidean space $\mathbb R^K$.
Geometrically, since $\sum_{k=1}^K\pi_{ik}=1$ with $\pi_{ik}\geq 0$, we know that each $\U_{i,:}$ is a convex combination of $\U_{S_1, :},\dots,\U_{S_K, :}$, which are the embeddings of the $K$ types of pure subjects. This means in $\mathbb R^K$, all the subjects lie in a simplex (i.e., the generalization of a triangle or tetrahedron to higher dimensions) whose vertices are these $K$ types of pure subjects.
Note that $\U_i$ and $\U_{i'}$ overlap if they have the same membership score vectors $\bo\pi_i = \bo\pi_{i'}$.
Figure \ref{fig-simplex} gives an illustration of this simplex geometry on such embeddings in $\mathbb R^3$ with $K=3$.

{Similar simplex structure in the spectral domain was first discovered and used for estimation under the degree-corrected mixed membership network model \citep[][first posted on arXiv in 2017]{jin2023mixed}. Later, spectral approaches to estimating mixed memberships via exploiting the simplex structure are also used for related network models \citep{mao2021mm} and topic models \citep{ke2022svd}. Compared to network models where the data matrix is symmetric, the GoM model has an $N\times J$ asymmetric data matrix. Compared to topic models, the entries of the perturbation matrix $\R-\R_0$ in the GoM model independently follow Bernoulli distributions, whereas in topic models, the entries of the perturbation matrix follow multinomial distributions.}

It is worth noting that the expectation identifiability in Definition \ref{def: identifiability} is closely related to the uniqueness of non-negative matrix factorization \citep[NMF,][]{donoho2003does, hoyer2004non, berry2007algorithms}. NMF seeks to decompose a nonnegative matrix $\M\in \mathbb{R}^{m\times n}$ into $\M = \W\bH$, where both $\W\in\mathbb{R}^{m\times r}$ and $\bH\in\mathbb{R}^{r\times n}$ are non-negative matrices. An NMF is called separable if each column of $\W$ appears as a column of $\M$ \citep{donoho2003does}. If we write $\M=\R_0^\top, \W=\bT, \bH=\bPi^\top$, then the separability condition for this NMF aligns with Condition \ref{condition1}. It is also generally assumed that $\W$ is of full rank, otherwise $\bH$ typically cannot be uniquely determined \citep{Gillis2013}.  Theorem \ref{Theorem2} will show that when Condition \ref{condition1} holds, $\bT$ being full-rank suffices for GoM model identifiability.
We will also show that model identifiability still holds with certain relaxations on the rank of $\bT$. 
On another note, estimating an NMF usually involves direct manipulation of the original data matrix, which can be computationally inefficient when dealing with large datasets. In our approach, we employ an NMF algorithm  in \cite{Gillis2013} on the SVD of the data matrix instead of the data matrix itself to estimate GoM parameters. Since our procedure operates on the singular subspace with significantly lower dimension than the original data space, it yields lower computational cost compared to conventional NMF procedures. 



We next present our identifiability results for GoM models.
We first show that the pure-subject Condition \ref{condition1} is almost necessary for the identifiability of a GoM model.

\begin{theorem}\label{Theorem1}
    Suppose $\theta_{jk}\in (0,1)$ for all $j=1,\dots, J$ and $k=1,\dots, K$. If there is one extreme profile that does not have any pure subject, then the GoM model is not identifiable.
\end{theorem}

The proofs of the theorems are all deferred to the Appendix.
Theorem \ref{Theorem1} reveals the importance of Condition \ref{condition1} for identifiability.
In fact, later we will use this condition as a foundation for our estimation algorithm.
Our next theorem presents sufficient and almost necessary conditions for GoM models to be identifiable.

\begin{theorem}\label{Theorem2}
    Suppose $\bo\Pi$ satisfies Condition \ref{condition1}. 
    \begin{enumerate}
        \item[(a)] If rank$(\bT)=K$, then the GoM model is identifiable.
        \item[(b)] If rank$(\bT)=K-1$ and no column of $\bT$ is an affine combination of the other columns of $\bT$, then the GoM model is identifiable. {\normalfont(}An affine combination of vectors $\x_1,\dots, \x_n$ is defined as $\sum_{i=1}^na_i\x_i$ with $\sum_{i=1}^na_i=1$.{\normalfont )}
        \item[(c)] In any other case, if there exists a subject $i$ such that $\pi_{ik}>0$ for every $k=1,\dots, K$,
        then the GoM model is not identifiable.
    \end{enumerate}
\end{theorem}

The high-level proof idea of Theorem \ref{Theorem2} shares a similar spirit to Theorem 2.1 in \cite{mao2021mm}.
We next explain and interpret the three settings in Theorem \ref{Theorem2}.
According to part (a), if the $K$ item parameter vectors $\bo\theta_1,\ldots,\bo\theta_K$ (i.e., the $K$ columns of $\bo\Theta$) are linearly independent, then the GoM model is identifiable under Condition \ref{condition1}. In part (b),
for identifiability to hold when $rank(\bT)=K-1$, any item parameter vector $\bo\theta_k$ cannot be written as an affine combination of the remaining vectors $\{\bo\theta_{k'}:\; k'\neq k\}$. This is a weaker requirement on $\bo\Theta$ compared to part (a). 
Part (c) states that if the conditions in parts (a) or (b) do not hold and if there exists a \emph{completely mixed subject} that partially belongs to all profiles (i.e., $\pi_{ik}>0$ for all $k$), then the model is not identifiable.
Part (c) also shows that the sufficient identifiability conditions in (a) and (b) are close to being necessary, because the existence of a completely mixed subject is a very mild assumption. 
We next further give three toy examples with $K=3$ and $J=4$ to illustrate the conditions in Theorem \ref{Theorem2}.

\begin{example}\label{Example1}
Consider
$$
\bT=
\begin{pmatrix}
0.2 & 0.8 & 0.8\\
0.2 & 0.8 & 0.2\\
0.8 & 0.2 & 0.8\\
0.8 & 0.2 & 0.2
\end{pmatrix}.
$$
It is easy to verify that $rank(\bT)=K=3$. This case falls into scenario (a) in Theorem \ref{Theorem2}, so a GoM model parameterized by $(\bPi, \bT)$ is identifiable if $\bPi$ satisfies Condition \ref{condition1}.
\end{example}

\begin{example}\label{Example2}
Consider
$$
\bT=
\begin{pmatrix}
0.2 & 0.8 & 0.8\\
0.2 & 0.8 & 0.8\\
0.8 & 0.2 & 0.8\\
0.8 & 0.2 & 0.8
\end{pmatrix}.
$$
Now $rank(\bT)=K-1=2$ since the third column of $\bT$ is a linear combination of the first two columns. However, there is no column of $\bT$ that is an affine combination of the other columns of $\bT$. This case falls into scenario (b) in Theorem \ref{Theorem2}, so a GoM model parameterized by $(\bPi, \bT)$ is identifiable if $\bPi$ satisfies Condition \ref{condition1}.
\end{example}

\begin{example}\label{Example3}
Consider
$$
\bT=
\begin{pmatrix}
0.2 & 0.8 & 0.5\\
0.2 & 0.8 & 0.5\\
0.8 & 0.2 & 0.5\\
0.8 & 0.2 & 0.5
\end{pmatrix}.
$$
In this case, $rank(\bT)=K-1=2$, and the third column of $\bT$ is an affine combination of the first two columns. This case falls into scenario (c) in Theorem \ref{Theorem2}, so if there exists a subject that partially belongs to all $K$ profiles, then the GoM model is not identifiable.
\end{example}

\section{SVD-based Spectral Estimation Method and Its Consistency} \label{sec: algorithm}

\subsection{Estimation Algorithm}
In the literature of GoM model estimation, the most prevailing approaches are perhaps Bayesian inferences using Markov chain Monte Carlo (MCMC) algorithms such as Gibbs and Metropolis-Hastings sampling \citep{erosheva2002grade, erosheva2007aoas, gu2023dimension}.
However, MCMC is time-consuming and typically not computationally efficient.
As \cite{borsboom2016kinds} points out, despite their usefulness, GoM models are somewhat underrepresented in psychometric applications due to the lack of readily accessible statistical software. 
Recently, the R package \texttt{sirt} \citep{robitzsch2022package} provides a joint maximum likelihood (JML) algorithm to fit GoM models. This algorithm implements the Lagrange multiplier method proposed in \cite{erosheva2002grade} and solves the optimization problem in a gradient descent fashion. 
Although this JML algorithm is computationally more efficient compared to MCMC algorithms, it is still not scalable to very large-scale response data due to its iterative manner. Therefore, it is of interest to develop a non-iterative estimation method suitable to analyze modern datasets with a large number of items and subjects.

We next propose a fast SVD-based spectral method to estimate GoM models.
Recall that Proposition \ref{prop1} establishes the expressions for $\bPi, \bT$ in \eqref{eq-pi_lemma1} and \eqref{eq-theta_lemma1}.
In practice, since $\S$ is not known, we propose to estimate it using a vertex-hunting technique called \emph{sequential projection algorithm} \citep[SPA;][]{araujo2001successive, Gillis2013}. As stated in Proposition \ref{prop1}, Condition \ref{condition1} induces a simplex geometry on the row vectors of $\U$ and the simplex vertices correspond to the pure subjects $\S$.
To locate $K$ vertices for any input matrix $\U\in\mathbb{R}^{N\times K}$ that has such a simplex structure, SPA first finds the subject with the maximum row norm in $\U$. That is, the first vertex index is 
$$\hat S_1=\argmax_{1\leq i\leq N} \|\U_{i,:}\|_2.$$  
Here and after we use $\|\x\|_2=\sqrt{\x^\top\x}$ to denote the $\ell_2$ norm of any vector $\x$. Since the $l_2$ norm of any convex combination of the vertices is at most the maximum $l_2$ norm of the vertices, this step is guaranteed to return one of the vertices of the simplex.
SPA then projects all the remaining row vectors $\{\U_{i,:}:\; i\neq \hat{S}_1\}$ onto the subspace that is orthogonal to $\U_{\hat{S}_1,:}$. Mathematically, denote $\mathbf{v}_1:=\U_{\hat{S}_1,:}/\|\U_{\hat{S}_1,:}\|_2$ as the scaled vector with unit norm, then the projected vectors are the rows of the matrix $\U(\mathbf{I}_K - \mathbf{v}_1\mathbf{v}_1^\top)$, where $\mathbf{I}_K - \mathbf{v}_1\mathbf{v}_1^\top$ is a $K\times K$ projection matrix. In the second step, SPA finds the second vertex index as the one that has the maximum norm among the projected row vectors,
$$
\hat S_2 = \argmax_{i\neq \hat S_1} \|\U_{i,:}(\mathbf{I}_K - \mathbf{v}_1\mathbf{v}_1^\top)\|_2.
$$
The above procedure of first finding the row with the maximum projected norm and then projecting the remaining rows onto the orthogonal space is iteratively employed until finding all $K$ vertex indices. 
Sequentially, for each $k=1,\ldots,K-1$, define a unit-norm vector $\mathbf{v}_k = \U_{\hat{S}_k,:}/\|\U_{\hat{S}_k,:}\|_2$, then the $(k+1)$-th vertex index is estimated as
$$
\hat S_{k+1} = \argmax_{i\not\in \{\hat S_1, \ldots,\hat S_k\}} \|\U_{i,:}(\mathbf{I}_K - \mathbf{v}_1\mathbf{v}_1^\top)\cdots (\mathbf{I}_K - \mathbf{v}_k\mathbf{v}_k^\top)\|_2.
$$
Here 
the projection matrices $(\mathbf{I}_K - \mathbf{v}_1\mathbf{v}_1^\top)$, $\dots$, $(\mathbf{I}_K - \mathbf{v}_k\mathbf{v}_k^\top)$ sequentially project the rows of $\U$ to the orthogonal spaces of those already found vertices $\U_{\hat S_1,:},\ldots, \U_{\hat S_k,:}$.
This SPA procedure can be intuitively understood by visually inspecting the toy example in Figure \ref{fig-simplex}. Since $\U_{1,:}, \ldots, \U_{N,:}$ lie in a triangle in Figure \ref{fig-simplex}, it is not hard to see that the vector with the largest norm should be one of the three vertices $\U_{S_1,:}, \U_{S_2,:}, \U_{S_3,:}$, say $\U_{S_3,:}$. Furthermore, after projecting the remaining $\U_{i,:}$ onto the space orthogonal to $\U_{S_3,:}$, the maximum norm of the projected vectors should correspond to $i=S_1$ or $S_2$. This observation intuitively justifies that SPA can find the correct set of pure subjects given a simplex structure.
With the estimated pure subjects $\hat\S$, $\bo\Pi$ and  $\bo\Theta$ can be subsequently obtained via \eqref{eq-pi_lemma1} and \eqref{eq-theta_lemma1}.

The above estimation procedure is based on the assumption that $\R_0$ is known. 
In practice, we only have access to the binary random data matrix $\R$ whose expectation is $\R_0$.
Fortunately, it is known that for a large-dimensional random matrix with a low-rank expectation, the top-$K$ SVD of the random matrix and the SVD of its expectation are close \citep[e.g., see][]{chen2021spectral}. 
This nontrivial theoretical result is our key insight and motivation to
consider the top-$K$ SVD of $\R$ as a surrogate for that of $\R_0$:
\begin{equation}
    \R\approx\hat{\U}\hat{\bS}\hat{\V}^\top,
\end{equation}
where $\hat{\bS}$ is a $K\times K$ diagonal matrix  collecting the $K$ largest singular values of $\R$, and $\hat{\U}_{N\times K}$, $\hat{\V}_{J\times K}$ collect the corresponding left and right singular vectors with $\hat{\U}^{\top}\hat{\U}=\hat{\V}^{\top}\hat{\V}=\mathbf{I}_K$. 
Specifically, \cite{chen2021spectral} proved that the difference between $\U$ and $\hat\U$ up to a rotation is small when $N$ and $J$ are large with respect to $K$.
Therefore, since Proposition \ref{prop1} shows that the population row vectors $\U_{1,:},\dots, \U_{N,:}$ form a simplex structure, the empirical row vectors $\{\hat{\U}_{i,:}\}_{i=1}^N$ are expected to form a noisy simplex cloud distributed around the population simplex. We call this noisy cloud the empirical simplex. 
For an illustration of the population and the empirical simplex, see Figure \ref{fig: sim_simplex} in Section \ref{sec: simulation}.

In order to recover the population simplex from the empirical one, 
we use a pruning step similar to that in \cite{mao2021mm} to reduce noise. We summarize this pruning procedure in Algorithm \ref{algorithm1}. 
The high-level idea behind pruning is that if  $\hat\U_{i,:}$ has a large norm but very few close neighbors, then it is likely to be outside of the population simplex and hence should be pruned (i.e., removed) before performing SPA to achieve higher accuracy in vertex hunting. 
More specifically, Algorithm \ref{algorithm1} first calculates the norm of each row of $\hat{\U}$ (line \textsc{1-3}) and identifies the vectors with norms in the upper $q$-quantile (line \textsc{4}). 
The larger $q$ is, the more such points are found. Then for each vector found, its average distance $x_i$ to its $r$ nearest neighbors is calculated (line \textsc{5-8}). Finally, the subjects to be pruned are those whose $x_i$ belong to the upper $e$-quantile of all the $x_i$'s (line \textsc{9}). 
A larger value of $e$ indicates that a larger proportion of the points will be pruned.
According to our preliminary simulations, we observe that the estimation results are not very sensitive to these tuning parameters $r,q,e$. 
After pruning, we can use SPA to hunt for the $K$ vertices of the pruned empirical simplex to obtain $\hat\S$, and then estimate $\bPi$ and $\bT$.

\begin{algorithm}[bt!]
    \centering
    \caption{Prune}\label{algorithm1}
    \begin{algorithmic}[1]
        \Require Empirical top-$K$ left singular matrix $\hat{\U}$, the number of nearest neighbors $r$, two quantiles $q,e\in(0,1)$.
        \Ensure Set of subject indices $\hat{\bP}$ to be pruned (i.e., removed)
        
        \For{$i\in 1,\dots, N$}
        \State $l_i=\|\hat{\U}_{i,:}\|_2$
        \EndFor
        \State $\hat{\bP}_0=\{i:l_i\ge \text{upper-$q$ quantile of }\{l_i: 1\le i\le N\}\}$
        \For{$i\in \hat{\bP}_0$}
        \State $\mathbf d_i=\{\text{the distances from } \hat{\U}_i \text{ to its } r \text{ nearest neighbors}\}$
        \State $x_i=\text{average}(\mathbf d_i)$
        \EndFor
        \State $\hat{\bP}=\{i: x_i\ge \text{upper-$e$ quantile of }\{x_i: i\in\hat{\bP}_0\}\}$
    \end{algorithmic}
\end{algorithm}

\begin{algorithm}[bt!]
    \centering
    \caption{GoM Estimation by Sequential Projection Algorithm with Pruning}\label{algorithm2}
    \begin{algorithmic}[1]
        \Require Binary response matrix $\R\in\{0,1\}^{N\times J}$, number of extreme profiles $K$, threshold parameter $\epsilon$, pruning parameters $r,q,e$.
        \Ensure Estimated $\hat{\S}, \hat{\bT}, \hat{\bPi}$
        \State Get the top $K$ singular value decomposition of $\R$ as $\hat{\U}\hat{\bS}\hat{\V}^\top$
        \State $\hat{\bP}$ = Prune$(\hat{\U}, r, q, e)$
        \State $\Y = \hat{\U}$
        \For{$k \in \{1,\dots,K\}$}                    
            \State $\hat{S}_k=\argmax (\{\|\Y_{i,:}\|_2:\; i\in[N] \backslash \hat{\bP}\})$
            \State $\mathbf{u} = \Y_{\hat{S}_k,:} / \|\Y_{\hat{S}_k,:}\|_2$
            \State $\Y = \Y (\I_K - \mathbf{u}\mathbf{u}^\top)$
        \EndFor
        \State $\tilde{\bPi}=\hat{\U}(\hat{\U}_{\hat{\S},:})^{-1}$
        \State $\hat{\bPi}=\text{diag}(\tilde{\bPi}_+ \mathbf{1}_K)^{-1} \tilde\bPi_+$
        \State $\tilde{\bT}=\hat{\V}\hat{\bS}\hat{\U}^{\top} \hat {\bPi} (\hat{\bo \Pi}^{\top}\hat{\bo \Pi})^{-1}
        $
         \State Write $\hat{\bT}$ as 
    $(\hat{\theta}_{j,k})$, where $\hat{\theta}_{j,k}=\begin{cases}
        \tilde{\theta}_{j,k} & \text{if } \epsilon\le \tilde{\theta}_{j,k}\le 1-\epsilon\\
        \epsilon & \text{if } \tilde{\theta}_{j,k}<\epsilon\\
        1-\epsilon & \text{if } \tilde{\theta}_{j,k}>1-\epsilon
    \end{cases}
    $

\noindent
($\epsilon\geq 0$ can be set to zero. In the numerical studies we choose $\epsilon = 0.001$ to be consistent and comparable with the default setting in the JML function in the R package \texttt{sirt}.)
\end{algorithmic}
\end{algorithm}

Algorithm \ref{algorithm2} summarizes our proposed method of estimating parameters $(\bPi,\bT)$ based on SPA with the pruning step.
We first introduce some notation. The index of the element in $\x$ that has the maximum value is denoted by $\argmax(\x)$. For any matrix $\A=(a_{ij})$, denote by $\A_+=(\max\{a_{ij}, 0\})$ the matrix that retains the nonnegative values of $\A$ and sets any negative values to zero.
Denote by $diag(\x)$ the diagonal matrix whose diagonals are the entries of the vector $\x$. If $\S_2$ is a subvector of $\S_1$, denote by $\S_1\backslash\S_2$ the complement of vector $\S_2$ in vector $\S_1$.
For a positive integer $M$, denote $[M] = \{1,\ldots,M\}$.
After obtaining the index vector of the pruned subjects $\hat{\bP}$ (which is a subvector of $(1,2,\ldots,N)$) from Algorithm \ref{algorithm1} (line \textsc{2}), 
we use SPA on the pruned matrix $\hat{\U}_{[N]\setminus\hat{\bP}, :}$ to obtain the estimated pure subject index vector $\hat{\S}$ (line \textsc{3-8}).
Once this is achieved, $\bPi$ and $\bT$ can be estimated by modifying \eqref{eq-pi_lemma1} and \eqref{eq-theta_lemma1} in Proposition \ref{prop1}. 
We first calculate
\begin{equation}\label{eq-pi_estimate_real}
    \tilde{\bo\Pi} = \hat{\U} \left(\hat{\U}_{\hat{\S},:}\right)^{-1}
\end{equation}
based on \eqref{eq-pi_lemma1}.
The $\tilde{\bo\Pi}$ obtained above does not necessarily fall into the parameter domain for $\bo\Pi$. Therefore, we first truncate all the entries of $\tilde{\bo\Pi}$ to be nonnegative, and then re-normalize each row to sum to one (line \textsc{10}). Based on $\hat\bPi$, we can also estimate $\bT$ by
\begin{equation} \label{eq-theta_estimate_real}
    \tilde{\bo \Theta}=\hat{\V}\hat{\bS}\hat{\U}^{\top} \hat {\bPi} (\hat{\bo \Pi}^{\top}\hat{\bo \Pi})^{-1}
\end{equation}
according to \eqref{eq-theta_lemma1}.
Our proposed method can be viewed as a method of moments. Equations \eqref{eq-pi_estimate_real} and \eqref{eq-theta_estimate_real} are based on the first moment of the response matrix $\R$, where we equate the low-rank structure of the population first-moment matrix $\R_0$ with the observed first-moment matrix $\R$.
Lastly, we truncate $\tilde\bT$ to be between $[\epsilon, 1-\epsilon]$ and obtain the final estimator $\hat\bT$ (line \textsc{12}).
When $\epsilon=0$, this truncation ensures that entries of $\hat\bT$ lie in the parameter domain $[0,1]$. 
In the numerical studies, we choose $\epsilon = 0.001$ to be consistent and comparable with the default setting in the JML function in the R package \texttt{sirt}.

\begin{remark}
There are two possible ways to estimate $\bo\Theta$ according to \eqref{eq-theta_lemma1}. The first one is defining $\hat\bT$ as the truncated version of
$$\tilde\bT=\hat\V\hat\bS\hat\U_{\hat\S,:}^\top,$$
which only uses the information of $\hat{\U}$ corresponding to the $K$ pure subjects indexed by $\hat{\S}$. Another approach is estimating $\bT$ via \eqref{eq-theta_estimate_real},
which uses information from all of the $N$ subjects. 
The latter approach is expected to give more stable estimates. Our preliminary simulations also justify that using \eqref{eq-theta_estimate_real} indeed gives higher estimation accuracy, so we choose to estimate $\bT$ that way.
\end{remark}

\subsection{Estimation Consistency}
In this subsection, we prove that our spectral method guarantees estimation consistency of both the individual membership scores $\bPi$ and the item parameters $\bT$.
We consider the double-asymptotic regime where both the number of subjects $N$ and the number of items $J$ grow to infinity and $K$ is fixed. 
At the high level, since our estimators are functions of the empirical SVD, we will prove consistency by leveraging singular subspace perturbation theory \citep{chen2021spectral} that quantifies the discrepancy between the SVD of $\R$ and that of the low-rank expectation $\R_0$.

Before stating the theorem, we introduce some notations. Write $f(n)\lesssim g(n)$ if there exists a constant $c > 0$ such that $|f(n)|\le c|g(n)|$ holds for all sufficiently large $n$, and write $f(n) \succsim g(n)$ if there exists a constant $c > 0$ such that $|f(n)| \ge c |g(n)|$ holds for all sufficiently large $n$. For any matrix $\A$, denote its $k$th largest singular value as $\sigma_k(\A)$, and define its condition number $\kappa(\A)$ as the ratio of its largest singular value to its smallest singular value.
For any $m\times n$ matrix  $\A=(a_{ij})\in\mathbb{R}^{m\times n}$, denote its Fronenius norm by $\|\A\|_F=\sqrt{\sum_{i=1}^m\sum_{j=1}^n a_{ij}^2}$.

\begin{condition}
\label{condition2}
$\kappa(\bPi)\lesssim 1, \kappa(\bT)\lesssim 1$, $\sigma_K(\bPi) \succsim \sqrt{N}$, and $\sigma_K(\bT) \succsim \sqrt{J}$.
\end{condition}

Condition \ref{condition2} is a reasonable and mild assumption on the parameter matrices. As a simple example, it is not hard to verify that if $\bPi$ and $\bT$ both consist of many identity submatrices $\I_K$ vertically stacked together, then Condition \ref{condition2} is satisfied.
The following theorem establishes the consistency of estimating both $\bPi$ and $\bT$ using our spectral estimator.

\begin{theorem}\label{thm: consistency}
Consider $\tilde{\bPi}=\hat\U\hat\U_{\hat{\S},:}^{-1}$, $\tilde \bT = \hat\V\hat{\bS} \hat\U^\top_{\hat\S,:}$.
Assume that Conditions \ref{condition1} and \ref{condition2} hold. If $N, J \to\infty$ with ${N}/{J^2}\to 0$ and  ${J}/{N^2}\to 0$, then we have
\begin{equation}
\frac{1}{\sqrt{NK}} \|\tilde{\bPi}-\bPi\bP\|_F \stackrel{P}{\to} 0, 
\quad
 \frac{1}{\sqrt{JK}} \|\tilde{\bT}\bP-\bT\|_F
\stackrel{P}{\to} 0,
\end{equation}
where the notation $\stackrel{P}{\to}$ means convergence in probability, and $\mathbf P$ is a $K\times K$ permutation matrix which has exactly one entry of ``1'' in each row/column and zero entries elsewhere.
\end{theorem}

Theorem \ref{thm: consistency} implies that
\begin{align*}
    \frac{1}{\sqrt{NK}} \|\tilde{\bPi}-\bPi\bP\|_F &= \sqrt{\frac{1}{NK} \sum_{i=1}^N \sum_{k=1}^K (\tilde\pi_{ik} - \pi_{i,\phi(k)})^2} ~\stackrel{P}{\to} 0;
    \\[3mm]
    \frac{1}{\sqrt{JK}} \|\tilde{\bT}-\bPi\bT\|_F &= \sqrt{\frac{1}{JK} \sum_{j=1}^J \sum_{k=1}^K (\tilde\theta_{jk} - \theta_{j,\phi(k)})^2} ~\stackrel{P}{\to} 0,
\end{align*}
where $\phi: \{1,\ldots,K\}\to \{1,\ldots,K\}$ is a permutation map determined by the $K\times K$ permutation matrix $\mathbf P$.
These results mean that as $N,J\to\infty$, the average squared estimation error across all entries in the mixed membership score matrix $\bPi$ and that across all entries in the item parameter matrix $\bT$ both converge to zero in probability.
This double-asymptotic regime with both $N$ and $J$ going to infinity and this consistency notion in scaled Frobenius norm are similar to those considered in the joint MLE approach to item factor analysis in \cite{chen2019joint} and \cite{chen2020structured}. 
To the best of our knowledge, this is the first time that consistency results are established for GoM models in this modern regime.

\section{Simulation Studies} \label{sec: simulation}

\subsection{Evaluating the Proposed Method}\label{simulation1}
We carry out simulation studies to evaluate the accuracy and computational efficiency of our new method. 
We consider $K\in\{3, 8\}$, $N\in\{200, 1000, 2000, 3000, 4000, 5000\}$, and $J=N/5$, 
which correspond to large-scale and high-dimensional data scenarios. Such simulation regimes share a similar spirit with those in \cite{zhang2020notesvd}, which proposed and evaluated an SVD-based approach for item factor analysis.
In each simulation setting, we generate 100 independent replicates. Within each setting and each replication, the rows of $\bPi$ are independently simulated from the Dirichlet$(\boldsymbol{\alpha})$ distribution with $\boldsymbol{\alpha}$ equal to the $K$-dimensional all-one vector, and the first $K$ rows of $\bPi$ are set to the identity matrix $\mathbf{I}_K$ in order to satisfy the pure-subject Condition \ref{condition1}. The entries in $\bT$ are independently simulated from the uniform distribution on $[0,1]$. 
Our preliminary simulations suggest setting the tuning parameters in the pruning Algorithm \ref{algorithm1} to $r=10$, $q=0.4$, $e=0.2$ with $8\%$ subjects removed before SPA, so that the pruned empirical simplex is adequately close to the population simplex.
We use the above setting throughout all the simulations and the real data analysis unless otherwise specified. It turns out the performance of our method is robust to these tuning parameter values and not much tuning is needed in practice.

\begin{figure}[h!]
    \centering
    \includegraphics[width=0.6\textwidth]{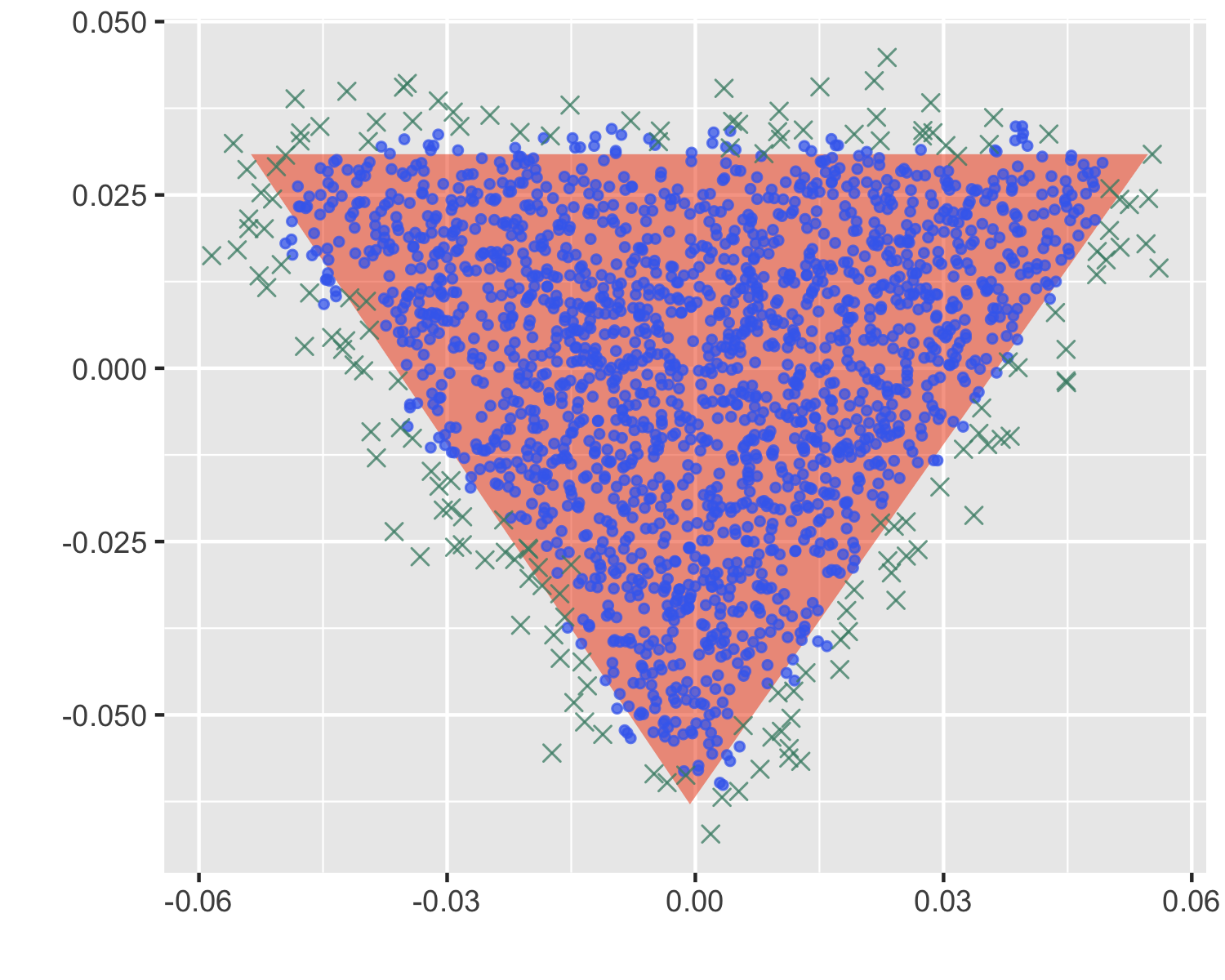}
    \caption{Row vectors of $\U$ and $\hat\U$ projected to $\mathbb R^2$ in the simulation setting with $N=2000$ and $K=3$. The red-shaded area is the population simplex, the green crosses are the removed subjects in pruning, and the blue dots form the empirical simplex retained after pruning.}
    \label{fig: sim_simplex}
\end{figure}

To illustrate the simplex geometry of the population and empirical left singular matrices $\U$ (from the SVD of $\R_0$) and $\hat\U$ (from the top-$K$ SVD of $\R$), we plot $\U_{i,:}$ and $\hat\U_{i,:}$ with $N=2000$ and $K=3$ in Figure \ref{fig: sim_simplex}. All vectors are projected to  two dimensions for better visualization of the simplex (i.e., triangle) structure. In Figure \ref{fig: sim_simplex}, the red-shaded area corresponds to the population simplex, the green crosses are the removed subjects selected by the pruning Algorithm \ref{algorithm1}, and the blue dots form the empirical simplex obtained after pruning. As we can see, the empirical row vectors $\hat\U_{i,:}$ approximately form a simplex cloud around the population simplex. After pruning out the noisy vectors, the resulting empirical simplex is close to the population one. This fact not only illustrates the effectiveness of the pruning procedure in Algorithm \ref{algorithm1}, but also confirms the usefulness of our notion of expectation identifiability by showing the close proximity of $\U$ and $\hat\U$. 
The resemblance between the scatter plots of the rows of $\U$ and $\hat\U$ implies that the simplex geometry of $\U$ holds approximately for $\hat\U$, and hence justifies that SPA can be applied to $\hat\U$ to estimate $\bPi$ and $\bT$.

We compare the performance of our proposed method with the \texttt{JML} algorithm (Joint Maximum Likelihood) in the R package \texttt{sirt}, because the latter is currently considered as  the most efficient estimation method for GoM models. We follow the default settings of \texttt{JML} with the maximum iteration number set as 600, the global parameter convergence criterion as 0.001, the maximum change in relative deviance as 0.001, and the minimum value of $\pi_{ik}$ and $\theta_{jk}$ as 0.001. We measure the parameter estimation error by the mean absolute error (MAE).
That is, the error between the estimated $\hat{\bo\Pi}$ and the ground truth $\bo\Pi$  for each replicate is quantified by the mean absolute bias (and similarly for $\bT$):
\begin{equation}\notag
    l(\bo\Pi, \hat{\bo\Pi}) := \frac{1}{NK}\sum_{i=1}^N\sum_{k=1}^K\vert \pi_{ik} - \hat{\pi}_{ik} \vert,\quad
    l(\bo\Theta, \hat{\bo\Theta}) := \frac{1}{JK}\sum_{j=1}^J\sum_{k=1}^K\vert \theta_{jk} - \hat{\theta}_{jk} \vert.
\end{equation}



\begin{figure}[h!]
    \centering
    \includegraphics[width=0.88\textwidth]{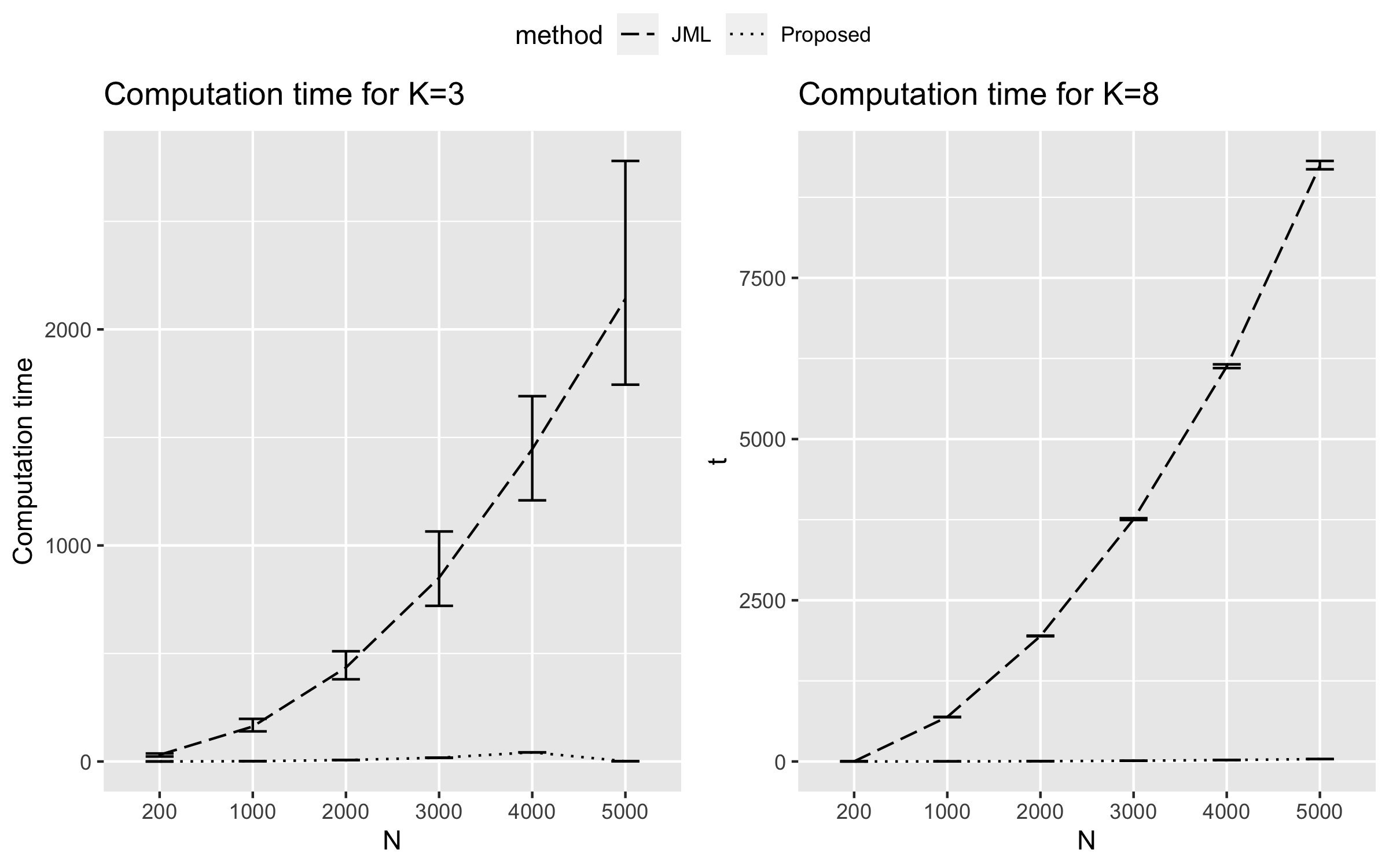}
    \caption{Computation time for $K=3$ (left) and $K=8$ (right) in simulations. For each simulation setting, we show the median, 25\% quantile, and 75\% quantile of the computation time for the 100 replications. 
    }
    \label{fig: sim_time}
\end{figure}
\begin{figure}[h!]
    \centering
    \includegraphics[width=0.9\textwidth]{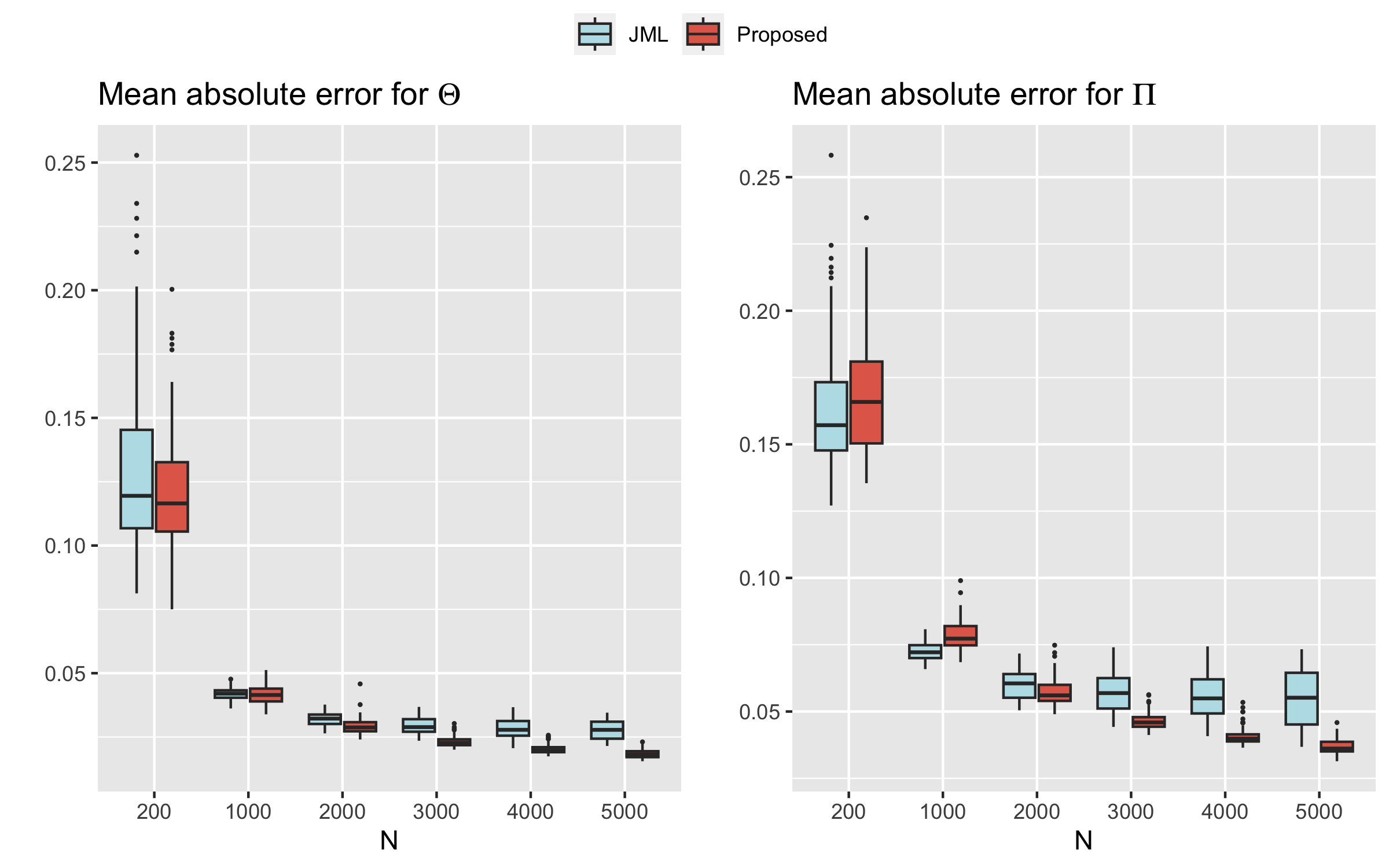}
    \caption{Simulation results of estimation error for $K=3$. The boxplots represent the 
    mean
    absolute error for $\bT$ (left) and $\bPi$ (right) versus the sample size $N$.}
    \label{fig: sim_err}
\end{figure}
\begin{figure}[h!]
    \centering
    \includegraphics[width=0.9\textwidth]{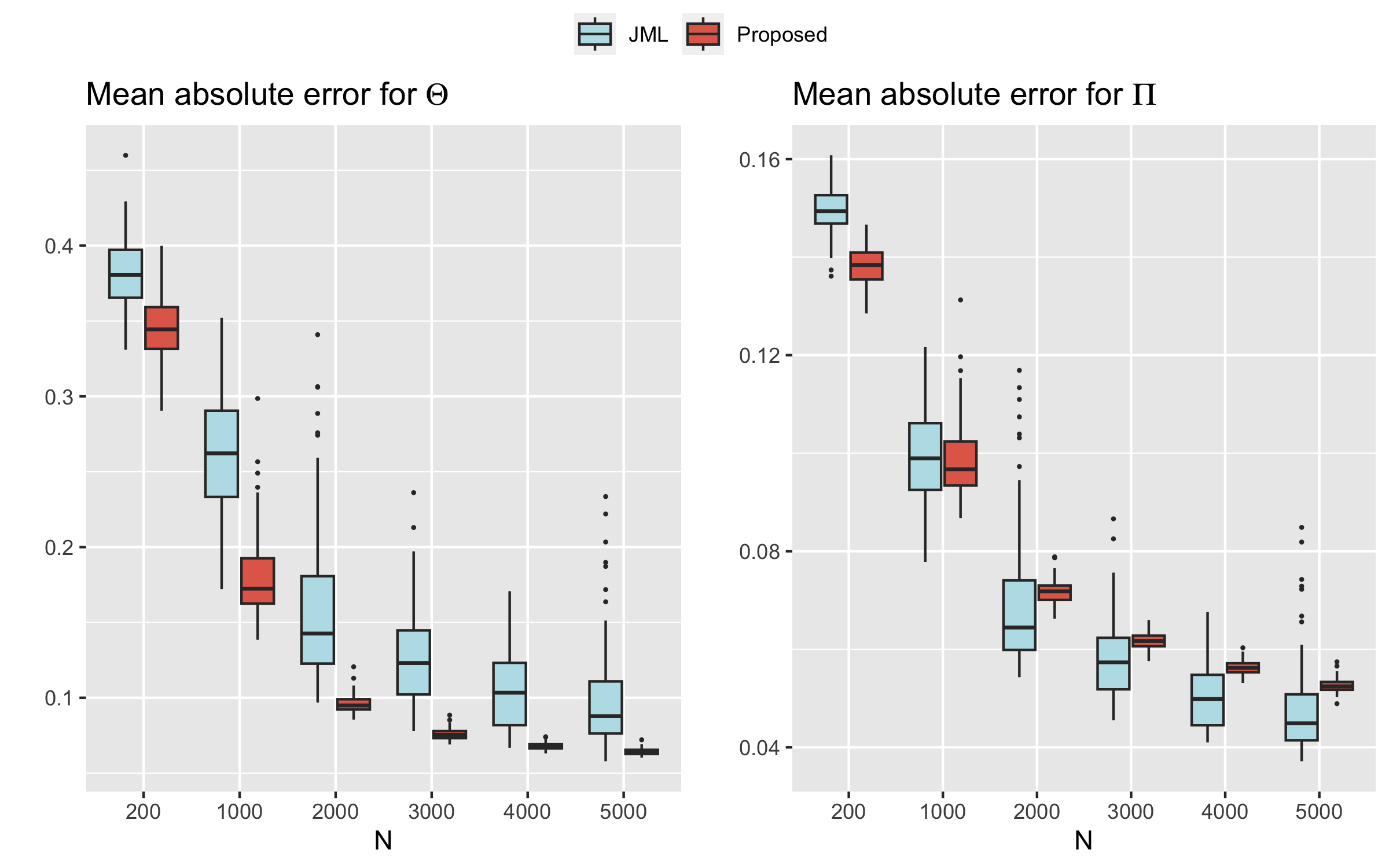}
    \caption{Simulation results of estimation error for $K=8$. The boxplots represent the mean absolute error for $\bT$ (left) and $\bPi$ (right) versus the sample size $N$.}
    \label{fig: sim_err_K=8}
\end{figure}

We present the comparisons between our spectral method and JML in terms of computation time and estimation error in Table \ref{table-time} and Figures \ref{fig: sim_time}, \ref{fig: sim_err}, \ref{fig: sim_err_K=8}. As shown in Figure \ref{fig: sim_time}, the computation time for both methods increases as the sample size $N$ and the number of items $J$ increase. Notably, JML takes significantly more time than our proposed method, especially when $N$ and $J$ are large. For example, when $N=5000$, $J=1000$, $K=8$, it takes about 3 hours for JML to reach the maximum iteration number for each replication, while it takes less than 40 seconds on average for our proposed method.
Table \ref{table-time} further records the mean computational time in seconds across replications for JML and the proposed method.
Moreover, we observe that JML is not able to converge in almost all replications when $K=8$, including when the sample size is as small as $N=200$.
In summary, when the number of extreme latent profiles $K$ is large enough, JML not only takes a long time to run but also is unable to reach convergence given the default convergence criterion. 

\begin{table}[h!] 
\centering
\small
\begin{tabular}{llllllll}
\toprule
$K$ & Method & $N=200$ & $N=1000$ & $N=2000$ & $N=3000$ & $N=4000$ & $N=5000$ \\
\midrule
\multirow{2}{*}{$K=3$} & JML & 30.3 & 170.9 & 473.9 & 923.4 & 1528.4 & 2363.9 \\
\cmidrule(lr){2-8} 
       & Proposed & 7.5e-2 & 1.5 & 6.8 & 17.5 & 42.3 & 10.6 \\
\midrule
\multirow{2}{*}{$K=8$} & JML & 1.7 & 691.5 & 1957.5 & 3761.7 & 6134.3 & 9251.7 \\
\cmidrule(lr){2-8} 
       & Proposed & 6.4e-2 & 1.3 & 5.0 & 11.6 & 23.0 & 37.2 \\
\bottomrule
\end{tabular}
\caption{Table of average computational time in seconds across replications for JML and the proposed spectral method for $K=3$ and $K=8$.}
\label{table-time}
\end{table}

The huge computational advantage of the proposed method does not come at the cost of degraded estimation accuracy. Figures \ref{fig: sim_err} and \ref{fig: sim_err_K=8} show that the estimation error decreases as the sample size $N$ and the number of items $J$ increase for both methods. When the sample size is large enough ($N\ge 2000$) for $K=3$, our proposed method gives more accurate estimation on average compared to JML. 
For $K=8$, our proposed method yields higher estimation accuracy for $\bT$ for all sample sizes on average. When $N\ge 2000$, the estimation accuracy of $\bPi$ is slightly worse but comparable to JML.
We point out that due to their non-iterative nature, SVD or eigendecomposition-based methods typically tend to give worse estimation compared to iterative methods that aim to find the MLE; for example, see the comparison of an SVD-based method and a joint MLE method for item factor analysis in \cite{zhang2020notesvd}. However, it turns out that given a fixed computational resource (i.e., the default maximum iteration number in the \texttt{sirt} package), the iterative method JML can give worse estimation accuracy than our spectral method.

We also compare our spectral method with the Gibbs sampling method for the GoM model. The parameters in the Gibbs sampler are initialized from their prior distributions. The number of burn-in samples is set to 5000. We take the average of 2000 samples after the burn-in phase as the Gibbs sampling estimates. Since Gibbs sampling is an MCMC algorithm, the computation can take a long time when the sample size $N$ and number of items $J$ are large. Therefore, we only consider $N=200, 1000, 2000$, and $K=3$. We compare the computation time and estimation accuracy of our proposed method, JML, and Gibbs sampling. The results are summarized in Tables \ref{table1} and \ref{table2}. Compared to Gibbs sampling, our proposed method is approximately 10,000 times faster when $N=2000$ and $K=3$. In terms of the estimation accuracy, when the sample size is small, the Gibbs sampling has better accuracy.
When the sample size is large enough, the three methods are comparable in terms of estimation accuracy. These simulation results  justify that our method is more suitable for large-scale and high-dimensional datasets.

\begin{table}[h!]
\centering
\begin{tabular}{c c c c} 
 \hline
  & $N=200$ & $N=1000$ & $N=2000$ \\ [0.5ex] 
 \hline
 Proposed & 0.1 & 1.5 & 6.8 \\ 
 JML & 30.3 & 170.9 & 473.9 \\
 Gibbs & 566.7 & 13119.2 & 52255.8 \\
 \hline
\end{tabular}
\caption{Average computation time in seconds for each method and sample size in simulations when $K=3$.}
\label{table1}
\end{table}

\begin{table}[h!]
\centering
\begin{tabular}{c c c c c c c c} 
 \hline
 $\bT$ & $N=200$ & $N=1000$ & $N=2000$ & $\bPi$ & $N=200$ & $N=1000$ & $N=2000$ \\ [0.5ex] 
 \hline
 Proposed & 0.12 & 0.04 & 0.03 & Proposed & 0.17 & 0.08 & 0.06\\ 
 JML & 0.13 & 0.04 & 0.03 &  JML & 0.16 & 0.07 & 0.06\\
 Gibbs & 0.09 & 0.03 & 0.02 &  Gibbs & 0.13 & 0.07 & 0.05\\
 \hline
\end{tabular}
\caption{Average mean absolute error for $\bT$ and $\bPi$ for each method and sample size in simulations when $K=3$.}
\label{table2}
\end{table}

We also note that without the pure subject Condition \ref{condition1}, 
one can still obtain estimates from the JML or the Gibbs sampler by directly running their corresponding estimation algorithms.
However, 
simply running those algorithms for arbitrary data generated under the GoM model may give misleading results if the model is not identifiable, because identifiability is the prerequisite for any valid statistical inference.
In contrast, for our proposed spectral estimator, under the pure subject condition, we have established both identifiability (Theorem \ref{Theorem1}) and estimation consistency (Theorem \ref{thm: consistency}) for both the mixed membership scores $\bPi$ and the item parameters $\bT$.

\begin{figure}[h!]
    \centering
    \includegraphics[width=0.95\textwidth]{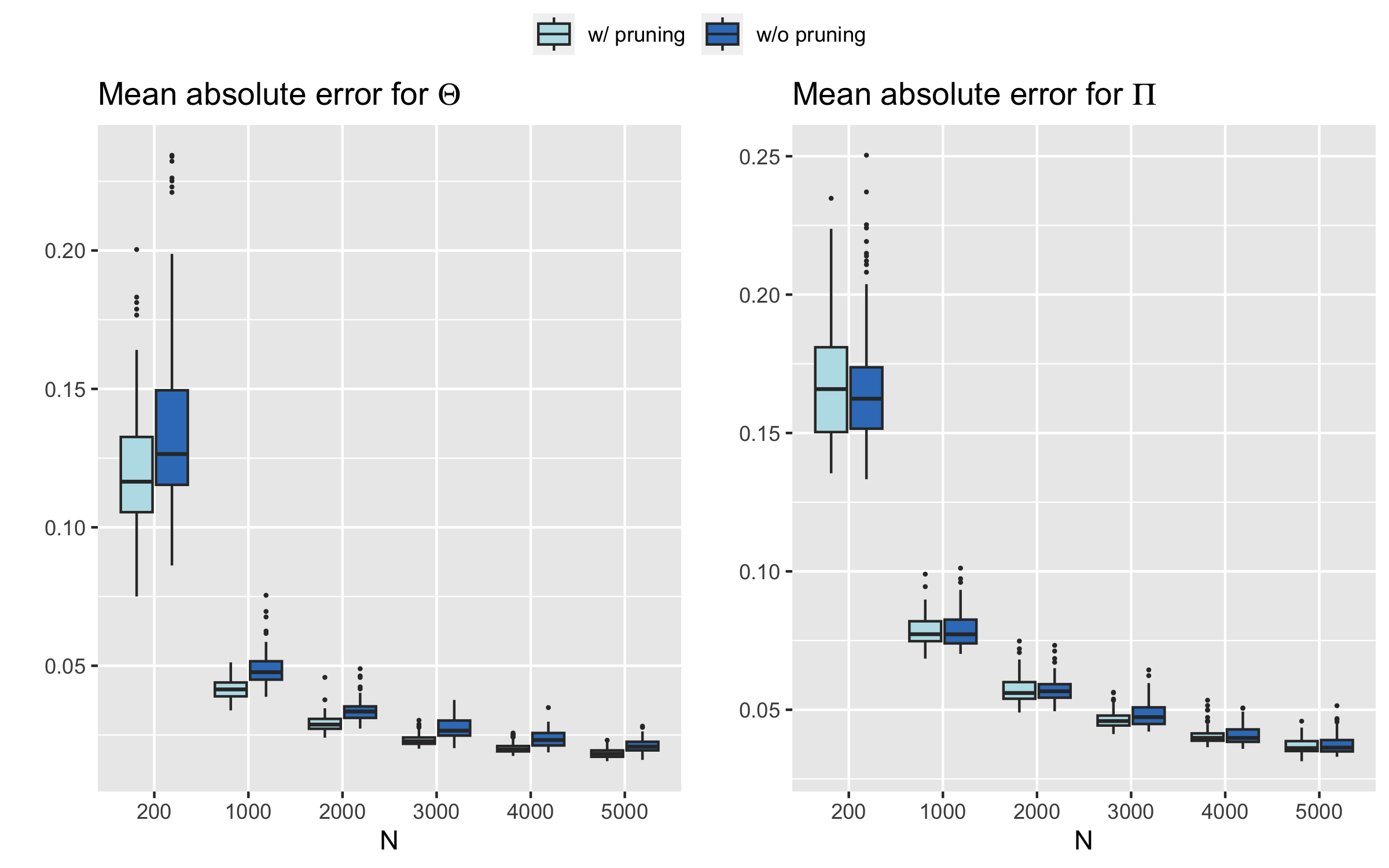}
    \caption{
    Comparison of estimation results with and without the pruning procedure when $K=3$.}
    \label{fig:pruning}
\end{figure}

We also compare the simulation results of our proposed method with and without the pruning step when $K=3$ to examine the effectiveness of pruning. The comparison of the estimation accuracy for the two approaches are summarized in Figure \ref{fig:pruning}. For the estimation of $\bT$, estimation with pruning is consistently better for all sample sizes. For the estimation of $\bPi$, estimation results are comparable for the two approaches. When sample size is large, pruning gives slightly better results.

To summarize, the above simulation results demonstrate the superior efficiency and accuracy of our proposed method to estimate GoM models.
Our method provides comparable estimation results compared with JML and Gibbs sampling, and is much more scalable to large datasets with many subjects and many items. 

\subsection{Verifying Identifiability}
\begin{figure}[h!]
    \centering
    \includegraphics[width=0.9\textwidth]{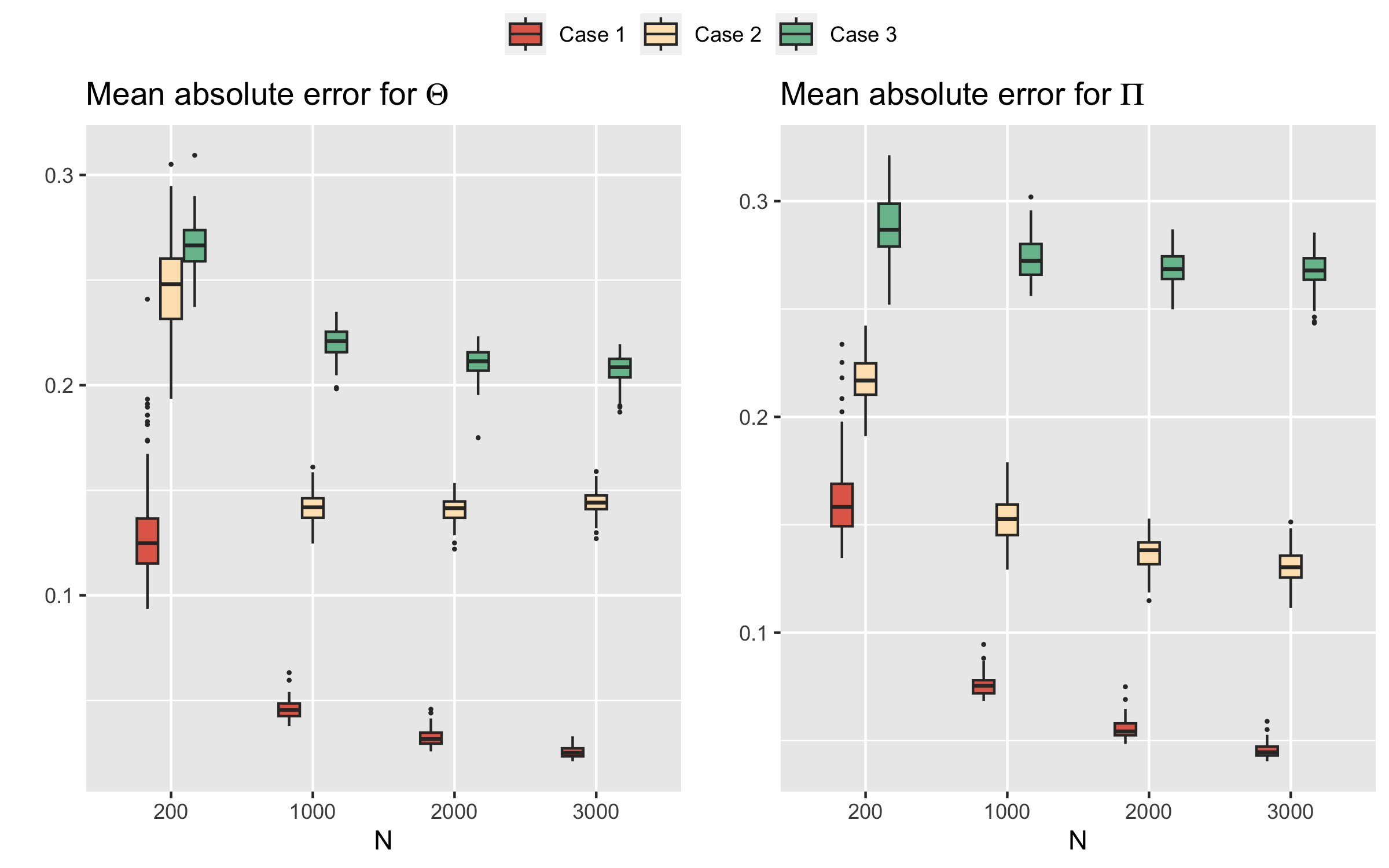}
    \caption{A simulation study verifying identifiability. Estimation errors for three different cases; see the concrete settings of Cases 1, 2, and 3 in the main text.
    The box plots represent the mean absolute error for $\bT$ (left) and $\bPi$ (right) versus the sample size $N$.}
    \label{fig: sim_err_ident}
\end{figure}

We also conduct another simulation study to verify the identifiability results. 
We consider three different cases with $K=3$, $N\in\{200,1000,2000,3000\}$, and $J=N/5$.
In Case 1, we set the ground truth $\bo\Theta$ by vertically concatenating copies of the identifiable $4\times 3$ $\bT$-matrix in Example \ref{Example1}, and the generation mechanism for $\bPi$ remains the same as in Section \ref{simulation1}. 
In Case 2, we use the same $\bo\Theta$ as in Case 1; while for $\bPi$, after generating rows of $\bPi'=(\pi'_{ik})$ from Dirichlet$(\mathbf 1)$, we truncate $\pi'_{ik}$ to be no less than $1/3$ and then re-normalize each row of it; after this operation the minimum entry in the resulting $\bo\Pi$ is 0.2.
Such a generated $\bo \Pi = (\pi_{ik})$ does not satisfy the pure-subject Condition \ref{condition1}.
In Case 3, we generate $\bPi$ using the same mechanism as in Section \ref{simulation1}, but generate $\bo\Theta$ whose rows are replicates of the vector $(0.8,~ 0.5,~ 0.2)$
so that $rank(\bo \Theta)=1$. Case 1 falls into part (a) in Theorem \ref{Theorem2} and is identifiable, whereas Cases 2 and 3 correspond to part (c) and are not identifiable. Figure \ref{fig: sim_err_ident} shows that the estimation errors in Cases 2 and 3 are significantly larger than those in Case 1. 
These results empirically verify our identifiability conclusions in Theorem \ref{Theorem2}.

\section{Real Data Example} \label{sec: real}
We illustrate our proposed method by applying it to a real-world personality test dataset, the Woodworth Psychoneurotic Inventory (WPI) dataset. WPI was designed by the United States Army during World War I to identify soldiers at risk for shell shock and is often credited as the first personality test. The WPI dataset can be downloaded from the Open Psychometrics Project website: \url{http://openpsychometrics.org/_rawdata/}. The dataset consists of binary yes/no  responses to $J=116$ items from 6019 subjects. We remove subjects with missing responses and only keep the subjects who are at least ten years old. This screening process leaves us with $N=3842$ subjects. 

\begin{figure}[h!]
    \centering
    \includegraphics[width=0.66\textwidth]{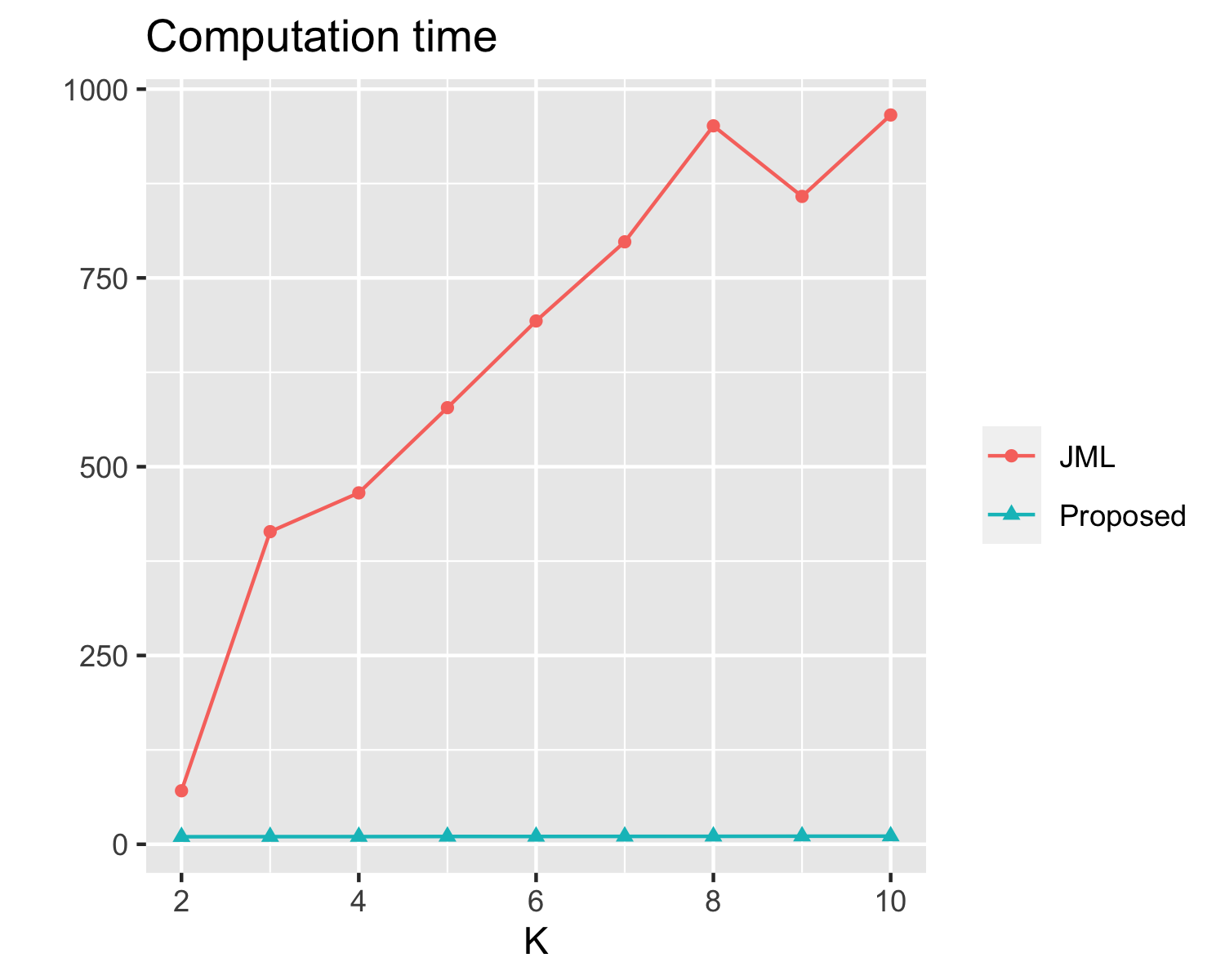}
    \caption{Computation time for the WPI dataset. The lines indicate the run time in seconds versus $K$ for JML and our spectral method. Note that for $K\ge 4$, the number of iterations in JML reaches the default maximum iteration number.}
    \label{fig: WPI_time}
\end{figure}

We apply both the JML method and our new spectral method to the WPI dataset to compare computation time. Figure \ref{fig: WPI_time} shows the computation time in seconds versus the number of extreme profiles $K$ for the two methods. For $K\ge 4$, the number of iterations for JML reaches the default maximum iteration number in the \texttt{sirt} package and does not converge. 
Similar to the simulations, our spectral method takes significantly less computation time compared to JML. This observation again confirms that the proposed method is scalable to real-world datasets with a large sample size and a relatively large number of items.

Choosing the number of extreme latent profiles $K$ is a nontrivial problem for GoM models in practice.
Available model selection techniques include 
the Akaike information criterion \citep[AIC; ][]{akaike1998information} and the Bayesian information criterion \cite[BIC;][]{schwarz1978estimating} for likelihood-based methods, and the deviance information criteria \cite[DIC;][]{spiegelhalter2002dic} for Bayesian MCMC methods. 
For factor analysis and principal component analysis, parallel analysis \citep{horn1965pa, dobriban2019deterministic} is one popular eigenvalue-based method to select the latent dimension. 
For the WPI dataset, when choosing $K=3$, we observe that the estimated $\hat\bT$ matrix has three well-separated column vectors, which imply a meaningful interpretation of the extreme profiles. When increasing $K$ to $4$, the columns of the estimated $\hat\bT$ are no longer that well-separated and interpretable. 
Moreover, choosing $K=3$ produces the least reconstruction error compared to $K=2$ or $4$. That is, $K=3$ leads to the smallest mean absolute error between $\hat\bPi\hat{\bT}^\top$ and the observed data matrix $\R$.
Since the goal of the current data analysis is mainly to illustrate the proposed spectral method, we next present and discuss the estimation results for $K=3$.
The important problem of how to select $K$ in GoM models in a principled manner is left as a future direction. 


 \begin{figure}[h!]
    \centering
    \includegraphics[width=0.9\textwidth]{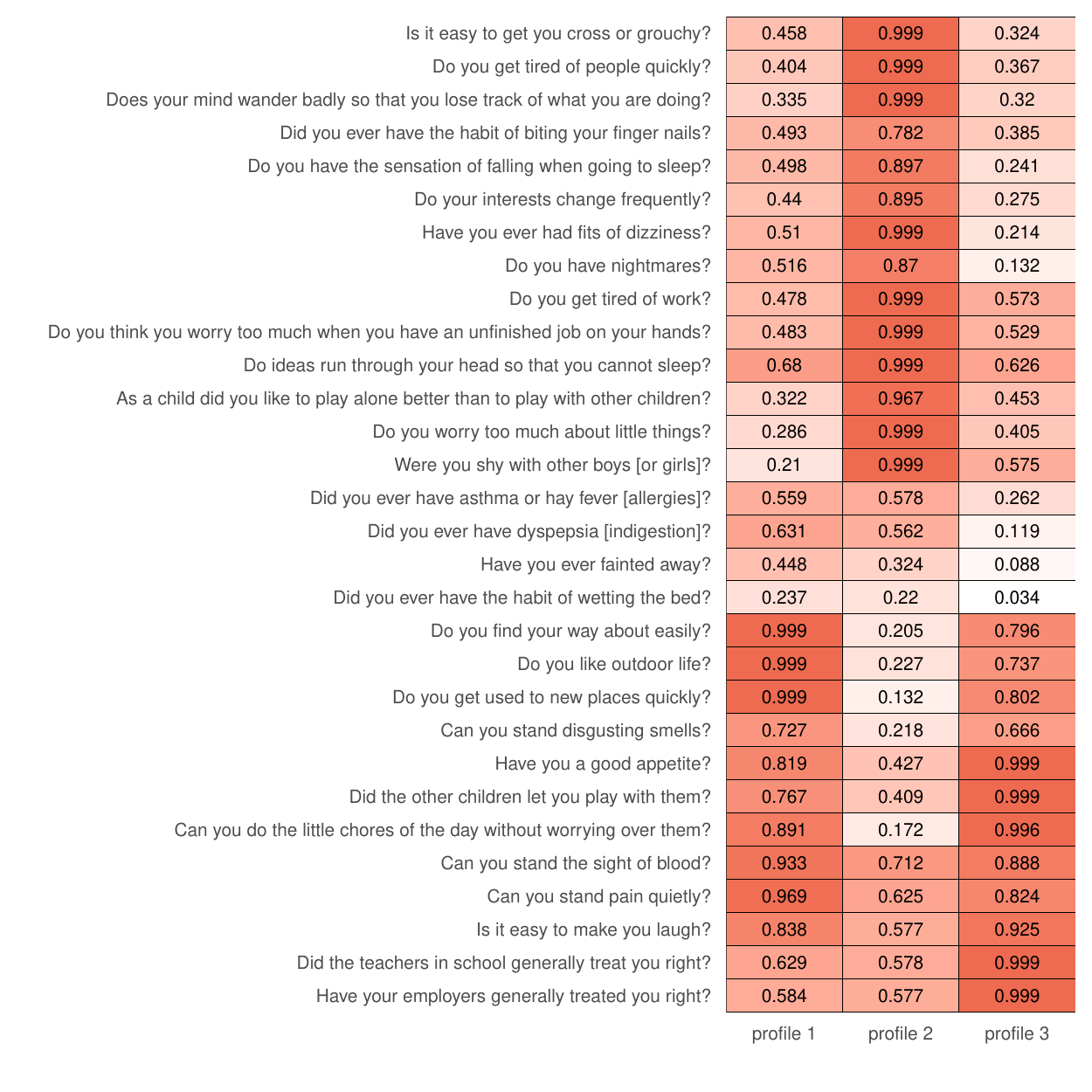}
    \caption{Heatmap of $\hat{\bT}$ of a subset of 30 WPI items. The values are the estimated probability of responding ``yes'' for each item given each extreme profile.}
    \label{fig: WPI_heat}
\end{figure}

We next take a closer look at the estimation result given by the proposed method for $K=3$.
In order to interpret each extreme latent profile, we present the heatmap of part of the estimated item parameter matrix $\hat{\bT}$ in Figure \ref{fig: WPI_heat}. Since the number of items $J=116$ is large, we only display a subset of the items for better visualization. More specifically, 30 out of 116 items are chosen in Figure \ref{fig: WPI_heat} based on the criterion that the chosen items have the largest variability in $(\hat\theta_{j,1}, \hat\theta_{j,2}, \hat\theta_{j,3})$.
Based on Figure \ref{fig: WPI_heat}, we interpret profile 1 as people who are physically unhealthy since they have higher probabilities of fainting, dyspepsia, and asthma or hay fever. People belonging to profile 2 tend to be socially passive since they are worrying, do not find their way easily, and get tired of things or people easily. Profile 3 on the other hand is identified as the healthy group.

\begin{figure}[h!]
    \centering
    \includegraphics[width=0.75\textwidth]{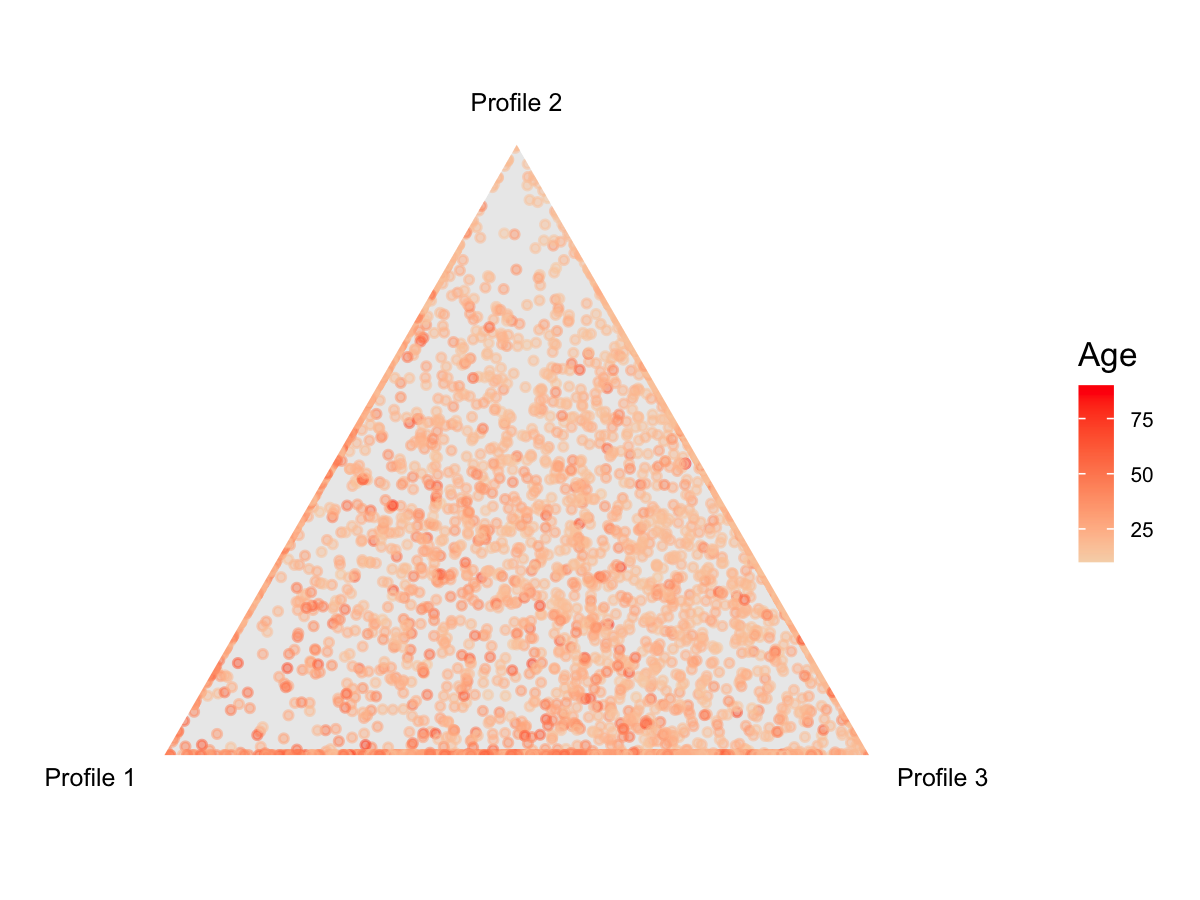}
    \caption{The barycentric plot of the estimated membership scores $\hat{\bPi}$ for WPI data, color-coded with the age covariate.}
    \label{fig: WPI_age}
\end{figure}

Figure \ref{fig: WPI_age} shows the ternary diagram of the estimated membership scores $\hat{\bPi}$ 
made with the R package \texttt{ggtern}.  The WPI dataset comes with the age information of each subject, and we color-code the subjects in Figure \ref{fig: WPI_age} according to their ages. The darker color represents older people while the lighter color represents younger people. 
Each dot represents a subject, and the location of the dot in the equilateral triangle depicts the three membership scores of this subject.
Specifically, dots with a large membership score on a profile are closer to the vertex of this profile. 
One can see that the pure-subject Condition \ref{condition1} is satisfied here since there are dots located almost exactly at each of the three vertices in Figure \ref{fig: WPI_age}.
This figure also reveals that darker dots are more gathered around the vertex of profile 1, which means that older people are more likely to belong to the extreme profile identified as physically unhealthy. Correspondingly, younger people are closer to profile 2 or 3. Recalling our earlier interpretation of the extreme profiles from Figure \ref{fig: WPI_heat}, these results are intuitively meaningful since older people tend to be less healthy compared with younger people. It is worth emphasizing that the age information is \emph{not used} in our estimation of the GoM model, but our method is able to generate interpretable results with respect to age. 

\section{Discussion} \label{sec: discussion}


In this paper, we have adopted a spectral approach to GoM analysis of multivariate binary responses. Under the notion of expectation identifiability, we have proposed sufficient conditions that are close to being necessary for GoM models to be identifiable. For estimation, we have proposed an efficient SVD-based spectral algorithm to estimate the subject-level and population-level parameters in the GoM model. 
Our spectral method has a huge computational advantage over Bayesian or likelihood-based methods and is scalable to large-scale and high-dimensional data. 
Simulation results demonstrate the superior efficiency and accuracy of our method, and also empirically corroborate our identifiability conclusions.
We hope this work provides a useful tool for psychometric researchers and practitioners by making GoM analysis less computationally daunting and statistically mysterious.

The expectation identifiability considered in this work is a suitable identifiability notion for large-scale and high-dimensional GoM models. 
A recent paper \cite{gu2023dimension} studied the {population identifiability} of a variant of the GoM model called the dimension-grouped mixed membership model. 
Generally speaking, population identifiability is a traditional notion of identifiability which aims at identifying the population parameters in the model but not the individual latent variables. In the context of the dimension-grouped GoM model, population identifiability in \cite{gu2023dimension} means identifying the item parameters $\bT$ and the distribution parameters for $\bPi$ (i.e., the Dirichlet distribution parameters $(\alpha_1,\ldots,\alpha_K)$, where each row in $\bPi$ is assumed to follow $\text{Dirichlet}(\alpha_1,\ldots,\alpha_K)$), but not identifying the entries in $\bPi$ directly. Such a traditional identifiability notion is more suitable for the low-dimensional case with small $J$ and large $N$, as considered in \cite{gu2023dimension}.
In contrast, in this work we are motivated by the large-scale and high-dimensional setting with both $N$ and $J$ going to infinity. In this setting, expectation identifiability that concerns both $\bT$ and $\bPi$ is a suitable notion to study, 
and we have also established the corresponding consistency result for $\bT$ and $\bPi$ when the pure subject condition for expectation identifiability is satisfied.

The pure subject Condition \ref{condition1} is a mild condition that is crucial to both our identifiability result and estimation procedure. It may be of interest to test whether this condition holds in practice.
Testing Condition \ref{condition1} is equivalent to testing whether a data cloud in a general $K$-dimensional space has a simplex structure, which is a non-trivial problem. 
When $K=3$, a visual inspection is plausible by plotting the row vectors of the left singular matrix and checking if the point cloud forms a triangle in the 3-dimensional space. However, when $K$ is larger than 3, visual inspection becomes infeasible.
Recently, a formal statistical procedure has been proposed by a working paper \cite{freyaldenhoven2023testability} to test the \emph{anchor word assumption} for topic models, which is another type of mixed membership models. 
The anchor word assumption requires that there exists at
least one anchor word for each topic, which is analogous to our pure subject Condition \ref{condition1}.
\cite{freyaldenhoven2023testability} considers the hypothesis testing of the existence of anchor words. 
They first characterize that for a matrix $\mathbf P$ that admits a low-rank factorization under the anchor word assumption, one can write $\mathbf P = \mathbf C\mathbf P$ for some matrix $\mathbf C$ that belongs to a certain set $\mathcal C_K$.
Based on this property, \cite{freyaldenhoven2023testability}  then construct a test statistic $T = \inf_{\mathbf C\in \mathcal{C}_K} \|\mathbf P - \mathbf C \mathbf P\|$ to test the null hypothesis that the anchor-word assumption holds. To achieve a level-$\alpha$ test, the null hypothesis will be rejected if $T$ is larger than the $(1-\alpha)\%$ quantile of the distribution of the test statistic under the null.
We conjecture that it might be possible to generalize this procedure to the GoM setting, and leave this direction as future work.

There are several additional research directions worth exploring in the future.
First, this work has focused on binary responses, while in practice it is also of interest to perform GoM analysis of polytomous responses, such as Likert-scale responses in psychological questionnaires \citep[see, e.g.][]{gu2023dimension}.
It would be desirable to extend our method to such multivariate polytomous data. Under a GoM model for polytomous data, a similar low-rank structure as \eqref{eq-model} should still exist for each category of the responses. Since our method is built upon the low-rank structure and the pure subject condition, we conjecture that exploiting such structures in polytomous data could still lead to nice spectral algorithms.
Second, it is worth considering developing a model that directly incorporates additional covariates into the GoM analysis. 
Our current method does not use additional covariates, such as the age information in the WPI dataset. Proposing a covariate-assisted GoM model and a corresponding spectral method may be methodologically interesting and practically useful.

\paragraph{Code Availability.} The R code implementing the proposed method is available at this link: \href{https://github.com/lscientific/spectral_GoM}{https://github.com/lscientific/spectral\_GoM}.

\begin{appendices}



\section{Proofs of the Identifiability Results}
\paragraph{Proof of Proposition~\ref{prop1}.} 
If we take the rows that correspond to $\S$ of both sides in the SVD \eqref{eq-svd} and use the fact that $\bPi_{\S,:}=\mathbf{I}_K$, then
\begin{equation}\label{eq-theta_U}
    \U_{\S,:}\bS\V^\top=[\R_0]_{\S,:}=\bPi_{\S,:} \bT^{\top}=\bT^{\top}.
\end{equation}
This gives an expression of $\bT$
\begin{equation}\label{eq-theta1}
    \bT = \V\bo\Sigma \U^\top_{\S,:}.
\end{equation}
Further note that 
\begin{equation}\label{eq-U}
    \U=\R_0\V\bS^{-1}=\bPi\bT^{\top}\V\bS^{-1}.
\end{equation}
If we plug \eqref{eq-theta_U} into \eqref{eq-U} and note that $\V$ have orthogonal columns, we have
\begin{equation}\label{eq-PiU}
    \U=\bPi\U_{\S,:}\bS\V^{\top}\V\bS^{-1}=\bPi\U_{\S,:}.
\end{equation}
Equation \eqref{eq-PiU} also tells us that $\U_{\S,:}$ must be full rank since both $\U$ and $\bPi$ have rank $K$. Therefore, we have an expression of $\bPi$:
\begin{equation}\notag
    \bo\Pi=\U (\U_{\S,:})^{-1}.
\end{equation}
On the other hand, based on the singular value decomposition $\U\bS\V^\top = \bo\Pi \bo\Theta^\top$, we can left multiply $(\bPi^{\top}\bPi)^{-1}\bo\Pi^\top$ with both hand sides to obtain
\begin{align} \notag
(\bPi^{\top}\bPi)^{-1}\bo\Pi^\top \U\bS\V^\top = \bo\Theta^\top
\\
\label{eq-theta_estimate}
\Longrightarrow\quad 
\bT=\V\bS\U^{\top}\bPi(\bPi^{\top}\bPi)^{-1}.
\end{align}
We next show that \eqref{eq-theta1} and \eqref{eq-theta_estimate} are equivalent. Since $\U^{\top}\U=\mathbf{I}_K$, \eqref{eq-PiU} leads to
\begin{equation}\notag
    \U_{\S,:}^\top\bPi^\top\bPi\U_{\S,:}=\mathbf{I}_K,
\end{equation}
which yields
\begin{equation}\notag
    (\bPi^{\top}\bPi)^{-1}=\U_{\S,:}\U^{\top}_{\S,:}.
\end{equation}
Plugging this equation into \eqref{eq-theta_estimate}, we have 
\begin{equation}\notag
\begin{split}
    \bT = & \V\bS\U^{\top}\bPi\U_{\S,:}\U^{\top}_{\S,:}\\
    = & \V\bS\U^{\top}\U (\U_{\S,:})^{-1}\U_{\S,:}\U^{\top}_{\S,:} \\
    = & \V\bS\U^{\top}_{\S,:}
\end{split}
\end{equation}
This shows the equivalence of \eqref{eq-theta1} and \eqref{eq-theta_estimate} and completes the proof of the proposition.
\qed

\paragraph{Proof of Theorem~\ref{Theorem1}.} 
Without loss of generality, assume that the first extreme latent profile does not have a pure subject. Then $\pi_{i1}\le 1-\delta,\ \forall i=1,\dots, N$ for some $\delta>0$. For each $0<\epsilon<\delta$, define a $K\times K$ matrix 
$$
\M_{\epsilon} =
\begin{bmatrix}
    1+(K-1)\epsilon^2 & -\epsilon^2\mathbf{1}_{K-1}^\top \\
    \mathbf{0}_{K-1} & \epsilon\mathbf{1}_{K-1}\mathbf{1}_{K-1}^\top+(1-(K-1)\epsilon)\mathbf{I}_{K-1}
\end{bmatrix}.
$$
We will show that $\tilde{\bPi}_{\epsilon}=\bPi\M_{\epsilon}$ and $\tilde{\bT}_{\epsilon}=\bT\M_{\epsilon}^{-1}$ form a valid parameter set. That is, each element of $\tilde{\bPi}_{\epsilon}$ and $\tilde{\bT}_{\epsilon}$ lies in $[0,1]$ and the rows of $\tilde{\bPi}_{\epsilon}$ sum to one. 
Since $\M_0=\mathbf{I}_K$ and by the continuity of matrix determinant, $\M_{\epsilon}$ is full rank when $\epsilon$ is small enough. Also notice that $\M_{\epsilon}\mathbf{1}_K=\mathbf{1}_K$. Therefore, $\tilde{\bPi}_{\epsilon}\mathbf{1}_K=\bPi\M_{\epsilon}\mathbf{1}_K=\bPi\mathbf{1}_K=\mathbf{1}_N$. For each $i=1,\dots, N$, $\tilde{\pi}_{i1}=\pi_{i1}(1+(K-1)\epsilon^2)\ge 0$. 
For any fixed $k=2,\dots,K$, $(\M_{\epsilon})_{kk} = 1 - (K-2)\epsilon$ and $(\M_{\epsilon})_{mk}=\epsilon$ for $m\neq k$. Thus when $\epsilon\leq 1/(K-1)$, we have $(\M_{\epsilon})_{mk}\ge \epsilon$ for any $m=1,\dots, K$. Therefore, the following inequalities hold for each $i=1,\dots, N$ and $k=2,\dots, K$:
\begin{align*}
    \tilde{\pi}_{ik}
    &=-\epsilon^2\pi_{i1}+\sum_{m=2}^K\pi_{im}(\M_{\epsilon})_{mk} {\ge} -\epsilon^2\pi_{i1}+\epsilon\sum_{m=2}^K\pi_{im}\\
    &\ge -\epsilon^2(1-\delta)+\epsilon(1-\pi_{i1})
    \ge -\epsilon^2(1-\delta) + \epsilon\delta\ge \epsilon\delta^2
    >0.
\end{align*}
Here we also used $\sum_{k=1}^K\pi_{ik}=1$, $\pi_{i1}\le 1-\delta$, and $\epsilon<\delta$. 

Further notice that 
$$
\M_{\epsilon}-\mathbf{I}_K=
\begin{bmatrix}
    \epsilon^2(K-1) & -\epsilon^2\mathbf{1}_{K-1}^\top \\
    \mathbf{0}_{K-1} & \epsilon\mathbf{1}_{K-1}\mathbf{1}_{K-1}^\top-\epsilon(K-1)\mathbf{I}_{K-1}
\end{bmatrix},
$$
which leads to $\|\M_{\epsilon}-\mathbf{I}_K\|_F\stackrel{\epsilon\rightarrow 0}{\longrightarrow}0$. Here $\|\A\|_F=\sqrt{\sum_{i=1}^m\sum_{j=1}^n a_{ij}^2}$ is the Fronenius norm of any matrix $\A=(a_{ij})\in\mathbb{R}^{m\times n}$. By the continuity of matrix inverse and Frobenius norm, $$\|\M_{\epsilon}^{-1}-\mathbf{I}_K\|_F\stackrel{\epsilon\rightarrow 0}{\longrightarrow}0.$$
Therefore, 
$$
\|\tilde{\bT}-\bT\|_F = \|\bT(\mathbf{I}_K-\M_{\epsilon}^{-1})\|_F\le \|\bT\|_2\|\mathbf{I}_K-\M_{\epsilon}^{-1}\|_F\stackrel{\epsilon\rightarrow 0}{\longrightarrow}0.
$$
Since all the elements of $\bT$ are strictly in $(0,1)$, the elements of $\tilde{\bT}$ must be in $[0,1]$ when $\epsilon$ is small enough. Also note that $\M_{\epsilon}$ is not a permutation matrix when $\epsilon>0$, thus the GoM model is not identifiable up to a permutation. This completes the proof of the theorem.
\qed

\paragraph{Proof of Theorem~\ref{Theorem2}.} 
Suppose $rank(\bT)=r\le K$. Now, consider SVD $\R_0=\U\bS \V^{\top}$ with $\U\in\mathbb{R}^{N\times r}, \V\in\mathbb{R}^{J\times r}, \bS\in\mathbb{R}^{r\times r}$. For simplicity, we continue to use the same notations $\U,\bS,\V$ here even though the matrix dimensions have changed. Without loss of generality, we can reorder the subjects and memberships so that $\bPi_{1:K,:}=\mathbf{I}_K$ from Assumption \ref{condition1}. According Proposition \ref{prop1},
\begin{equation}\label{Eq:11}
    \U=\bPi\U_{1:K,:}.
\end{equation}
Since $rank(\bPi)=K, rank(\U)=r$, we must have $\rank(\U_{1:K,:})=\rank(\U)=r$.

Suppose another set of parameters $(\tilde{\bPi},\tilde{\bT})$ yields the same $\R_0$ and we denote its corresponding pure subject index vector as $\tilde{\S}$ so that $\tilde{\bPi}_{\tilde{\S},:}=\mathbf{I}_K$. Similarly, we have
\begin{equation}\label{Eq:12}
    \U=\tilde{\bPi}\U_{\tilde{\S},:}.
\end{equation}
Taking the $\tilde{\S}$ rows of both sides of \eqref{Eq:11} and the first $K$ rows of both sides of \eqref{Eq:12} yields
\begin{equation}\notag
\label{Eq:13}
    \bPi_{\tilde{\S},:}\U_{1:K,:}=\U_{\tilde{\S},:},\ \U_{1:K,:}=\tilde{\bPi}_{1:K,:}\U_{\tilde{\S},:}.
\end{equation}
The above equation shows that $\U_{\tilde{\S},:}$ is in the convex hull created by the rows of $\U_{1:K,:}$, and $\U_{1:K,:}$ is in the convex hull created by the rows of $\U_{\tilde{\S},:}$. Therefore, there must exist a permutation matrix $\bP$ such that $\U_{\tilde{\S},:}=\bP\U_{1:K,:}$. Combining this fact with \eqref{Eq:11} and \eqref{Eq:12} leads to
\begin{equation}\label{Eq:14}
    (\bPi - \tilde{\bPi}\bP) \U_{1:K,:} = 0.
\end{equation}

\paragraph{Proof of part (a).}
For part (a), $r=K$ and $\U_{1:K,:}$ is full rank according to \eqref{Eq:11}. In this case, \eqref{Eq:14} directly leads to $\bPi=\tilde{\bPi}\bP$ and thus $\tilde{\bT}=\bT\bP^\top$.

Now generally consider $r<K$. By permuting the rows and columns of $\bT$, we can write
\begin{equation}\label{Eq:15}
    \bT = 
    \begin{bmatrix}
        \C & \C\W_1 \\
        \W_2^\top\C & \W_2^\top\C\W_1
    \end{bmatrix},
\end{equation}
where $\C\in\mathbb{R}^{r\times r}$ is full rank, $\W_1\in\mathbb{R}^{r\times (K-r)}$ and $\W_2\in\mathbb{R}^{r\times (J-r)}$. Now comparing the block columns of \eqref{Eq:15} and $\bT=\V\bS(\U_{1:K,:})^\top$ gives
\begin{equation}\label{Eq:16}
    \begin{split}
        \begin{bmatrix}
            \mathbf{I}_r \\
            \W_2^\top
        \end{bmatrix}\C & = \V\bS(\U_{1:r,:})^\top,\\
        \begin{bmatrix}
            \mathbf{I}_r \\
            \W_2^\top
        \end{bmatrix}\C\W_1 & = \V\bS(\U_{(r+1):N,:})^\top.
    \end{split}
\end{equation}
Since $\C$ is full rank, $\U_{1:r,:}$ has to also be full rank and \eqref{Eq:16} can be translated into 
$$\U_{(r+1):N,:}=\W_1^\top \U_{1:r,:}.$$ Therefore,
\begin{equation}\label{eq-24}
    \U_{1:K,:}=\begin{bmatrix}
        \mathbf{I}_r \\
        \W_1^\top
    \end{bmatrix}\U_{1:r,:}.
\end{equation}
By plugging the \eqref{eq-24} into \eqref{Eq:14} and again using the fact that $\U_{1:r,:}$ is full rank, we have
\begin{equation}\label{Eq:18}
    (\bPi-\tilde{\bPi}\bP) \begin{bmatrix}
        \mathbf{I}_r \\
        \W_1^\top
    \end{bmatrix} = \mathbf{0}.
\end{equation}

\paragraph{Proof of part (b).}
Denote $\A:=\bPi-\tilde{\bPi}\bP$. If $r=K-1$, $\W_1 = (W_{1,1}, \dots, W_{1,K-1})$ is a $(K-1)$-dimensional vector and \eqref{Eq:18} gives us
\begin{equation}\label{Eq:19}
    \A_{:,j} + W_{1,j}\A_{:,k}=0, \quad \forall j=1,\dots, K-1.
\end{equation}
Denote an $r$-dimensional column vector with all entries equal to one by  $\one_r$. 
Right multiplying $\mathbf{1}_r$ to both sides of \eqref{Eq:18} yields
$$\A\begin{bmatrix}
    \mathbf{1}_r \\
    \W_1^\top\mathbf{1}_r
\end{bmatrix} = 0.
$$
Also, note that both $\bPi$ and $\tilde{\bPi}\bP$ have row sums of 1. Hence,
\begin{equation}\notag
    \begin{split}
         \sum_{j=1}^{K}\A_{:,j} =\mathbf{0}_N,\\
        \sum_{j=1}^{K-1}\A_{:,j} + \W_1^\top\mathbf{1}_r\A_{:,K} =\mathbf{0}_N.
    \end{split}
\end{equation}
Taking the difference of the two equations above gives $(1-\W_1^\top\mathbf{1}_r) \A_{:,K} = \zero_N$.
If $\W_1^\top\mathbf{1}_r\neq 1$, then $\A_{:,K}$ has to be $\zero_N$, which implies $\A_{:,j}=\zero_N$ for all $j=1\dots, K-1$ according to \eqref{Eq:19}. Therefore, $\A=\bPi-\tilde{\bPi}\bP=\zero_N$, which leads to $\tilde{\bT}=\bT\bP^\top$.  

Note that using \eqref{eq-24} leads to
$$\bT^{\top}=\U_{1:K,:}\bS\V^\top=\begin{bmatrix}
        \mathbf{I}_{K-1} \\
        \W_1^\top
    \end{bmatrix}\U_{1:(K-1),:}\bS\V^\top = \begin{bmatrix}
        \mathbf{I}_{K-1} \\
        \W_1^\top
    \end{bmatrix}(\bT_{:,1:(K-1)})^\top.$$
Hence $\bT_{:,K}=\bT_{:,1:(K-1)}\W_1$. Therefore, the condition $\W_1^\top\mathbf{1}_r\neq 1$ is equivalent to the $K$-th column of $\bT$ not being an affine combination of the other columns.

\paragraph{Proof of part (c).}
Now consider the case of \emph{either} $r=K-1$ with $\W_1^\top\mathbf{1}_r=1$, \emph{or} $r<K-1$. Assume subject $m$ is completely mixed so that $\pi_{m,k}>0,\forall k=1,\dots, K$. Define
\begin{equation}\notag
    \tilde{\bpi}_i^\top=
    \begin{cases}
        \bpi_i^{\top} & \quad \text{if } i\neq m\\
        \bpi_m^{\top} + \epsilon\boldsymbol{\beta}^\top [-\W_1^\top ,~ \mathbf{I}_{K-r}]  & \quad \text{if } i=m
     \end{cases},
\end{equation}
where $\epsilon>0$ is small enough so that $\tilde{\bpi}_i\in(0,1)$, and $\boldsymbol{\beta}\in\mathbb R^{K-r}$ is such that $\boldsymbol{\beta}^\top(\mathbf{1}_{K-r}-\W_1^\top \mathbf{1}_{r})=0$. Note that such $\boldsymbol{\beta} \neq \zero$ always exists under the assumption in part (c), because if $r=K-1$ with $\W_1^\top\mathbf{1}_r=1$, then $\boldsymbol{\beta}^\top(\mathbf{1}_{K-r}-\W_1^\top \mathbf{1}_{r}) = \beta (1-1)=0$ holds for any $\beta \in \mathbb R$; if $r<K-1$, then $K-r\geq 2$ and $\bo\beta$ has dimension at least two, so the inner product equation $\boldsymbol{\beta}^\top(\mathbf{1}_{K-r}-\W_1^\top \mathbf{1}_{r})=0$ must have a nonzero solution $\bo\beta$. The constructed $\tilde{\bPi}$ have row sums of 1 by the construction of $\boldsymbol{\beta}$. Furthermore, $\tilde{\bPi}\U_{1:K,:}$ and $\bPi\U_{1:K,:}$ can only be different on the $m$-th row, and 
$$\tilde{\bpi}_m^\top \U_{1:K,:}=\bpi^{\top}_m\U_{1:K,:}+ \epsilon \boldsymbol{\beta}^\top[-\W_1^\top, ~ \mathbf{I}_{K-r}]  \begin{bmatrix}
    \mathbf{I}_r \\
    \W_1^\top
\end{bmatrix}\U_{1:r,:}=\bpi_m^{\top}\U_{1:K,:}.$$
Hence $\tilde{\bPi}\U_{1:K,:}=\bPi\U_{1:K,:}$. This gives us
\begin{equation}\notag
    \bPi\bT^{\top} = \bPi \U_{1:K,:}\bS\V^\top = \tilde{\bPi}\U_{1:K,:}\bS\V^\top=\tilde{\bPi}\bT^{\top}.
\end{equation}
We can see that $(\bPi,\bT)$ and $(\tilde{\bPi}, \bT)$ yield the same model but $\bPi\neq \tilde{\bPi}$. This completes the proof for part (c). 
\qed

\section{Proof of the Consistency Theorem \ref{thm: consistency}}
For any matrix $\A$ with SVD $\A=\U_{\A}\bS_{\A}\V_{\A}^\top$, define 
\begin{equation*}
    \text{sgn}(\A) := \U_{\A}\V_{\A}^\top.
\end{equation*}
According to Remark 4.1 in \cite{chen2021spectral}, for any two matrices $\A, \B\in\mathbb{R}^{n\times r}, r\le n$:
\begin{equation*}
    \text{sgn}(\A^\top\B) = \arg\min_{\bO \in\mathcal{O}^{r\times r}} \|\A\bO - \B\|,
\end{equation*}
where $\mathcal{O}^{r\times r}$ is the set of all orthonormal matrices of size $r\times r$. 
The $2$-to-$\infty$ norm of matrix $\A$ is defined as the maximum row $l_2$ norm, i.e., $\norm{\A}_{2,\infty}=\max_i \|\e_i^\top \A\|$.
Define $$r=\frac{\max\{N,J\}}{\min\{N,J\}}.$$

Under Condition \ref{condition2}, we have $\kappa(\R_0) = \frac{\sigma_1(\bPi\bT^\top)}{\sigma_K(\bPi\bT^\top)} \le \frac{\sigma_1(\bPi)\sigma_1(\bT)}{\sigma_K(\bPi)\sigma_K(\bT)} = \kappa(\bPi)\kappa(\bT)\lesssim 1$ and $\sigma_K(\R_0) \geq \sigma_K(\bPi) \sigma_K(\bT) \succsim \sqrt{NJ}$. 

\begin{lemma}
\label{lemma-bound}
Under Condition \ref{condition2}, if $N/J^2\rightarrow 0$ and $J/N^2\rightarrow 0$, then with probability at least $1-O((N+J)^{-5})$, one has 
\begin{align}\label{eq-bound-1}
    \|\hat\U -\U\cdot\text{sgn}(\U^\top\hat\U)\|_{2,\infty} &\lesssim \frac{\sqrt{r} + \sqrt{\log (N+J)}}{\sqrt{NJ}}
\\
\label{eq-bound-2}
    \|\hat\U\hat\bS\hat\V^\top - \U\bo\Sigma \V^\top\|_{\infty} &\lesssim \sqrt{\frac{r\log(N+J)}{\min\{N,J\}}}.
\end{align}
\end{lemma}
Here the infinity norm $\|\A\|_{\infty}$ for any matrix $\A$ is defined as the maximum absolute entry value.
We write the RHS of \eqref{eq-bound-1} as $\varepsilon$ and the RHS of \eqref{eq-bound-2} as $\eta$. 

\paragraph{Proof of Lemma \ref{lemma-bound}.}
We will use Theorem 4.4 in \cite{chen2021spectral} to prove the lemma and will verify the conditions of that theorem are satisfied.

Define the incoherence parameter $\mu := \max\left \{\frac{N\norm{\U}_{2,\infty}^2}{K}, \frac{J\norm{\V}_{2,\infty}^2}{K}\right\}$. Note that
\begin{equation*}
    \|\U\|_{2,\infty} \le \|\U_{\S,:}\|_{2,\infty} \leq \|\U_{\S,:}\| 
    = \frac{1}{\sigma_K(\bPi)} \lesssim \frac{1}{\sqrt{N}},
\end{equation*} 
since all rows of $\U$ are convex combinations of $\U_{\S,:}$. On the other hand,
\begin{align*}
    \|\V\|_{2,\infty} = & \|\bT \U^{-\top}_{\S,:}\bS^{-1}\|_{2,\infty} \le \|\bT\|_{2,\infty} \|\U^{-\top}_{\S,:}\bS^{-1}\| \\
    \le & \|\bT\|_{2,\infty} \|\U^{-1}_{\S,:}\| \cdot \frac{1}{\sigma_K(\bPi\bT^{\top})} 
    = \frac{\|\bT\|_{2,\infty}\sigma_1(\bPi)}{\sigma_K(\bPi\bT^{\top})} \\
    \le & \frac{\|\bT\|_{2,\infty}\kappa(\bPi)}{\sigma_K(\bT)} \leq \frac{\sqrt{K} \kappa(\bPi)}{\sigma_K(\bT)} \lesssim \frac{1}{\sqrt{J}}. 
\end{align*}
Therefore, $\mu \lesssim 1$. 

On the other hand, we will show that $\sqrt{\log(N+J)/\min\{N, J\}}\lesssim 1$. By the symmetry of $N$ and $J$, we assume $J\le N$ without loss of generality. Thus
\begin{align*}
    \sqrt{\frac{\log(N+J)}{\min\{N, J\}}} =\sqrt{\frac{\log(N+J)}{J}} \lesssim \sqrt{\frac{\log(J^2+J)}{J}} \to 0.
\end{align*}

Therefore, Assumption 4.2 in \cite{chen2021spectral} holds and \eqref{eq-bound-1} and \eqref{eq-bound-2} can be directly obtained from Theorem 4.4 in \cite{chen2021spectral}.
\qed

\begin{lemma}\label{lemma-spa}
Let Conditions \ref{condition1} and \ref{condition2} hold. Then, there exists a permutation matrix $\bP$ such that with probability at least $1-O((N+J)^{-5})$,
\begin{equation}
    \|\hat\U_{\hat\S,:} - \bP\U_{\S,:}\cdot\text{sgn}(\U^\top\hat\U)\| \lesssim \varepsilon.
\end{equation}
\end{lemma}
\paragraph{Proof of Lemma \ref{lemma-spa}.}
Using Proposition \ref{prop1}, we will apply Theorem 4 in \cite{klopp2021assigning} with $\tilde{\mathbf G}=\hat \U$, $\mathbf G=\U\cdot\text{sgn}(\U^\top\hat\U)$, $\mathbf W = \bPi$, $\Q=\U_{\S,:}\cdot\text{sgn}(\U^\top\hat\U)$, $\mathbf N=\hat\U-\U\text{sgn}(\U^\top\hat\U)$, $\mathbf N = \hat\U-\U\cdot\text{sgn}(\U^\top\hat\U)$. According to Lemma \ref{lemma-bound}, $\|\e_i^\top\mathbf N\|\le \varepsilon$ and $\varepsilon \lesssim \frac{\sqrt{r}+\sqrt{\log(N+J)}}{\sqrt{NJ}}$. On the other hand, $\sigma_K(\mathbf Q)=\sigma_K(\U_{\S,:})=\frac{1}{\sigma_1(\bPi)}\ge \frac{1}{\sqrt{N}}$ since $\U=\bPi\U_{\S,:}$ and $\sigma_1(\bPi) \leq \norm{\bPi}_F \leq \sqrt{N} \max_i\norm{\e_i^\top\bPi}_{2} \leq \sqrt{N}$. Therefore, $\varepsilon\le C_*\frac{\sigma_K(\U_{\S,:})}{K\sqrt{K}}$ for some $C_*>0$ small enough. Then we can use Theorem 4 in \cite{klopp2021assigning} to get
\begin{align*}
     \|\hat\U - \bP\U_{\S,:}\cdot \text{sgn}(\U^\top\hat\U)\| & \le C_0\sqrt{K}\kappa(\U_{\S,:}\cdot\text{sgn}(\U^\top\hat\U))\varepsilon\\
     & = C_0\sqrt{K}\kappa(\U_{\S,:})\varepsilon \stackrel{(i)}{=} C_0\sqrt{K}\kappa(\bPi)\varepsilon \\
     & \lesssim \varepsilon ~~ \text{ with probability at least $1-O((N+J)^{-5})$.}
\end{align*}
Here $(i)$ is because $\U=\bPi\U_{\S,:}$.
\qed

\paragraph{Proof of Theorem \ref{thm: consistency}.}
First show that $\hat\U_{\hat{\S},:}$ is not degenerate. 
By Weyl's inequality and Lemma \ref{lemma-spa}, with probability at least $1-O((N+J)^{-5})$, we have
\begin{align*}
    \sigma_K(\hat\U_{\hat{\S},:}) & \ge \sigma_K(\bP\U_{\S,:}\bO) - \|\hat\U_{\hat{\S},:} - \bP\U_{\S,:}\cdot\text{sgn}(\U^\top\hat\U)\|\\
    & \ge \sigma_K(\U_{\S,:}) - \|\hat\U_{\hat{\S},:} - \bP\U_{\S,:}\cdot\text{sgn}(\U^\top\hat\U)\|_F\\
    & \succsim \frac{1}{\sigma_1(\bPi)} - \varepsilon\\
    & \succsim \frac{1}{\sqrt{N}} - \frac{\sqrt{r}+\sqrt{\log(N+J)}}{\sqrt{NJ}}\\
    & \succsim \frac{1}{\sqrt{N}}
\end{align*}
when $N, J$ are large enough and $\frac{N}{J^2}$ converges to zero.
Therefore, $\hat\U_{\hat{\S},:}$ is invertible.

For the estimation of $\bPi$,
\begin{align*}
        \|\tilde{\bPi}-\bPi\bP\|_F = & \|\hat\U\hat\U_{\hat{\S},:}^{-1} - \U\U^{-1}_{\S,:}\bP\|_F\\
        \le & \underbrace{\|\hat\U(\hat\U_{\hat{\S},:}^{-1} - \text{sgn}(\U^\top\hat\U)^\top\U^{-1}_{\S,:}\bP)\|_F}_{I_1} + \underbrace{\|(\hat\U - \U \text{sgn}(\U^\top\hat\U))[\bP^{-1}\U_{\S,:}\text{sgn}(\U^\top\hat\U)]^{-1}\|_F}_{I_2}\\
        =: & I_1 + I_2.
\end{align*}
We will look at $I_1$ and $I_2$ separately.
\begin{align*}
    I_1 & = \|\hat\U(\hat\U_{\hat{\S},:}^{-1} - \text{sgn}(\U^\top\hat\U)^\top\U^{-1}_{\S,:}\bP^\top)\|_F \\
    & \le \|\hat\U_{\hat{\S},:}^{-1} - \text{sgn}(\U^\top\hat\U)^\top\U^{-1}_{\S,:}\bP^\top\|_F\\
    & \le \|\hat\U_{\hat{\S},:}^{-1}\| \|\text{sgn}(\U^\top\hat\U)^\top\U^{-1}_{\S,:}\bP^\top\|\|\hat\U_{\hat{\S},:} - \bP\U_{\S,:}\text{sgn}(\U^\top\hat\U)\|_F\\
    & \lesssim \sqrt{N} \cdot \sigma_1(\bPi) \cdot \varepsilon\\
    & \lesssim \sqrt{N}\cdot \frac{\sqrt{r}+\sqrt{\log(N+J)}}{\sqrt{J}} ~~ \text{with probability at least $1-O((N+J)^{-5})$;}
\end{align*}
and
\begin{align*}
    I_2 = & \|(\hat\U - \U \text{sgn}(\U^\top\hat\U))[\bP^{-1}\U_{\S,:}\text{sgn}(\U^\top\hat\U)]^{-1}\|_F\\
    \le & \|\hat\U - \U \text{sgn}(\U^\top\hat\U)\|_F \|\U_{\S,:}^{-1}\| \\
    \le & \sqrt{N} \cdot \varepsilon \cdot \sigma_1(\bPi) \\
    \lesssim & \sqrt{N}\cdot \frac{\sqrt{r}+\sqrt{\log(N+J)}}{\sqrt{J}}~~ \text{with probability at least $1-O((N+J)^{-5})$.}
\end{align*}
Therefore, with probability at least $1-O((N+J)^{-5})$,
\begin{align*}
    \frac{1}{\sqrt{NK}} \|\tilde{\bPi}-\bPi\bP\|_F \lesssim & \frac{\sqrt{r}+\sqrt{\log(N+J)}}{\sqrt{J}}\\
    = & \begin{cases}
        \frac{\sqrt{N}}{J} + \frac{\sqrt{\log(N+J)}}{\sqrt{J}} & \text{if } N > J, \\
        \frac{1}{\sqrt{N}} + \frac{\sqrt{\log(N+J)}}{\sqrt{J}} & \text{if } N \le J.
    \end{cases}
\end{align*}
Therefore, $\frac{1}{\sqrt{NK}} \|\tilde{\bPi}-\bPi\bP\|_F$ converges to zero in probability as $N, J\to\infty$ and $\frac{N}{J^2}\to 0$.

For the estimation of $\bT$,
\begin{align*}
    & \|\tilde{\bT}\bP-\bT\|_F \\
    = & \|\bP^\top\hat\U_{\hat\S,:}\hat\bS\hat\V^\top - \U_{\S,:}\bS\V^{\top}\|_F\\
    \le & \| (\bP^\top\hat\U_{\hat\S,:} - \U_{\S,:}\text{sgn}(\U^\top\hat\U))\hat\bS\hat\V^{\top}\|_F + \|(\U_{\S,:}\text{sgn}(\U^\top\hat\U) - \hat\U_{\S,:})\hat\bS\hat\V^\top\|_F  \\
    & + \|\hat\U_{\S,:} \hat\bS \hat\V^\top - \U_{\S,:}\bS\V^{\top}\|_F\\
    \le & \| \bP^\top\hat\U_{\hat\S,:} - \U_{\S,:}\text{sgn}(\U^\top\hat\U)\|_F\cdot \sigma_1(\R)\cdot \|\hat\V\| + \|\U_{\S,:}\text{sgn}(\U^\top\hat\U) - \hat\U_{\S,:}\|\cdot \sigma_1(\R)\cdot \|\hat\V\|\\
    & + \sqrt{KJ}\|\hat\U_{\S,:} \hat\bS \hat\V^\top - \U_{\S,:}\bS\V^{\top}\|_{\infty}\\
    \stackrel{(ii)}{\lesssim} & \varepsilon \cdot \sigma_1(\R_0) + \varepsilon \cdot (\sigma_1(\R) - \sigma_1(\R_0)) + \sqrt{KJ}\cdot \eta~~ \text{with probability at least $1-O((N+J)^{-5})$},
\end{align*}
where $(ii)$ results from Lemma \ref{lemma-spa}. 
By Weyl's inequality, $|\sigma_1(\R) - \sigma_1(\R_0)| \leq \norm{\R-\R_0}$, where $\R-\R_0$ is a mean-zero Bernoulli matrix. According to Eq (3.9) in \cite{chen2021spectral}, with probability at least $1-(N+J)^{-8}$, $$
\norm{\R-\R_0}\lesssim \sqrt{N+J} + \sqrt{\log(N+J)}.
$$
Furthermore, $\sigma_1(\R_0)\ge \sigma_K(\R_0)\succsim \sqrt{NJ}$ by Condition \ref{condition2}, thus we know that $\sigma_1(\R_0) \succsim |\sigma_1(\R) - \sigma_1(\R^*)|$ with probability at least $1-(N+J)^{-8}$. Therefore, with probability at least $1-O((N+J)^{-5})$, 
\begin{align*}
     \|\hat{\bT}\bP-\bT\|_F & \lesssim \varepsilon \cdot \sigma_1(\R_0) + \sqrt{KJ}\cdot \eta\\
     & \lesssim \frac{\sqrt{r} + \sqrt{\log (N+J)}}{\sqrt{NJ}} \cdot \sqrt{N} \cdot \sqrt{J} + \sqrt{J} \sqrt{\frac{r\log(N+J)}{\min\{N,J\}}}\\
     & = \sqrt{r} + \sqrt{\log (N+J)} + \sqrt{J} \sqrt{\frac{r\log(N+J)}{\min\{N,J\}}}.
\end{align*}
Thus
\begin{align*}
    \frac{1}{\sqrt{JK}} \|\hat{\bT}\bP-\bT\|_F & \lesssim \frac{\sqrt{r} + \sqrt{\log (N+J)}}{\sqrt{J}} + \sqrt{\frac{r\log(N+J)}{\min\{N,J\}}} \\
    & = \begin{cases}
        \frac{\sqrt{N}}{J}+ \frac{\sqrt{\log (N+J)}}{\sqrt{J}} + \frac{\sqrt{N\log(N+J)}}{J} & \text{if } N > J \\
        \frac{1}{\sqrt{N}} + \frac{\sqrt{\log (N+J)}}{\sqrt{J}} + \frac{\sqrt{J\log(N+J)}}{N} & \text{if } N \le J
    \end{cases}.
\end{align*}
Therefore,  $\frac{1}{\sqrt{JK}} \|\hat{\bT}\bP-\bT\|_F$ converges to zero in probability as $N, J\to\infty$ and $\frac{N}{J^2}, \frac{J}{N^2}\to 0$.
\qed
\end{appendices}

\spacingset{1}

\bibliographystyle{apalike}
\bibliography{ref}

\end{document}